\documentclass[10pt, draftcls, onecolumn, journal]{IEEEtran}
\usepackage{cite, graphicx,amssymb,color, slashbox, pict2e, multirow, rotating}
\usepackage{graphicx,amssymb}
\usepackage[tbtags]{amsmath}
\usepackage{times,amsmath}
\usepackage{subfig}
\usepackage{flushend}
\usepackage{amsmath}
\usepackage{epsfig}
\usepackage{amssymb}
\usepackage{hyperref}
\usepackage{color}
\usepackage{psfrag}
\usepackage{caption}
\usepackage{subfig}
\usepackage[usenames,dvipsnames]{xcolor}
\usepackage{textcomp}
\usepackage{array}
\usepackage{tikz}
\usetikzlibrary{matrix,decorations.pathreplacing}

\usepackage{tabularx,ragged2e}
\newcolumntype{C}{>{\Centering\arraybackslash}X} 

\usepackage{algorithmic}
\usepackage{algorithm}

\ifCLASSINFOpdf
\else
\fi

\begin{document}
%
\title{Network Utility Maximization based on Incentive Mechanism for Truthful Reporting of Local Information}


\author{\IEEEauthorblockN{Jie Gao, \IEEEmembership{Member, IEEE},   
		Lian Zhao, \IEEEmembership{Senior Member, IEEE}, 
		and Xuemin (Sherman) Shen, \IEEEmembership{Fellow, IEEE} 
	}

\thanks{
Copyright (c) 2015 IEEE. Personal use of this material is permitted. However, permission to use this material for any other purposes must be obtained from the IEEE by sending a request to pubs-permissions@ieee.org.

J. Gao and L. Zhao are with the Department of Electrical and Computer Engineering, Ryerson University, Toronto, ON, M5B 2K3, Canada (e-mail: \{j.gao, l5zhao\}@ryerson.ca).
 	
X. Shen is with the Department of Electrical and Computer Engineering, University of Waterloo, Waterloo, ON, N2L 3G1, Canada (e-mail: sshen@uwaterloo.ca). 
}
}


%


\maketitle

\begin{abstract}
Classic network utility maximization problems are usually solved assuming all information is available, implying that information not locally available is always truthfully reported. This may not be practical in all scenarios, especially in distributed/semi-distributed networks. In this paper, incentive for truthful reporting in network optimizations with local information is studied.  A novel general model for extending network utility maximization (NUM) problems to incorporate local information is proposed, which allows each user to choose its own objective locally and/or privately. Two specific problems, i.e., a user-centric problem (UCP) and a network-centric problem (NCP), are studied. In the UCP, a network center aims to maximize the collective benefit of all users, and truthful reporting from the users regarding their local information is necessary for finding the solution. We show that the widely-adopted dual pricing cannot guarantee truthful information reporting from a user unless the resource is over-supplied or the price is too high for this user to afford. In the NCP, the network center has its own objective and preferred solution, and incentive needs to be provided for the users to adopt the solution it prefers. Truthful reporting from users is necessary for the center to determine the incentives and achieve its solution. For two-user and multiuser cases, we propose two mechanisms to motivate truthful reporting from users while guaranteeing nonnegative utility gains for both the users and the center. A case study on underlay D2D communication illustrates the application of the UCP and NCP. Simulations are conducted for the D2D application to validate the analytical results and demonstrate the proposed mechanisms.

\end{abstract}


%
\IEEEpeerreviewmaketitle

\section{Introduction}


\subsection{Motivation}

Network utility maximization (NUM) problems have been studied in many applications including sensor networks, vehicular networks, cellular networks, and smart grids \cite{JYJ2013} - \cite{DT2016}. Most of these studies adopt an optimization perspective, which focuses on the properties of a problem such as its convexity and decomposability, e.g., \cite{DPalomar2006}. Meanwhile, it is generally assumed that information required for finding the solution is available to the corresponding entities (e.g., the nodes running the optimization algorithm) either locally or through truthful communications. These assumptions can be well-justified for networks with a central administrator, e.g., cellular networks, or those with simple capability-limited nodes, e.g., sensor networks. 

However, with the increasing popularity of heterogeneous networks and Internet of things (IoT) \cite{SZhang2016}\cite{YWu2017TMC}, more applications tend to be based on a distributed or a semi-distributed network structure \cite{YZhou2016TWC}\cite{KMalekshan2017TWC}. Meanwhile, network nodes have increasingly more computation, communication, and/or other (e.g., sensing) capabilities. As a result, a network node may simultaneously possess local information, computation capability for information processing and decision making, and an individual benefit-maximizing objective. In such scenarios, it can be argued that assuming the availability of local information while finding the solution to an optimization problem is less practical. After all, a node is not motivated to share its true information if it cannot expect a benefit by doing so. As a result, the solution that can be found from an optimization perspective may not be achieved due to the existence of individual objectives and local information. 

The above scenario requires an incentive mechanism to be adopted as it can motivate individual users to truthfully report their local information while finding the solution.

\subsection{Related Works}
Mechanism design is an analytical framework in the field of game theory. It considers the implementation of desirable system-wide solutions to problems when a solution depends on self-interested agents and the information for the agents to make decisions is dispersed and locally held \cite{RVohra2011}. 
Game theoretic perspectives for modeling decision making have been adopted in many research works, e.g.,  \cite{OSemiari2015}-\cite{NZhang2017}.  While game theory typically studies the outcome of a group of players interacting with each other, mechanism design concerns providing appropriate incentives for the players so that their interaction can yield a desirable outcome \cite{YN2009}, e.g., a preferred solution to a considered optimization problem. Mechanism design has been adopted in an increasing number of works in communications and networking, especially in the past five years. In \cite{PSamadi2012}, the classic Vickrey-Clarke-Groves (VCG) mechanism was adopted to maximize social welfare for a smart grid. Iosifidis \textit{et al.} proposed an iterative double auction mechanism for efficient mobile data offloading \cite{GIosifidis}. A tax mechanism based on the classic Groves mechanism was introduced in \cite{TT2013} to implement dual decomposition so that a user has a diminishing incentive to report falsely as the algorithm converges. Wu \textit{et al.} used a game theoretic design in dynamic spectrum access and proposed a pricing mechanism for users to adopt the social welfare optimizing solution \cite{YWU2011}. Jin \textit{et al.} considered incentive for cloudlet service provisioning in mobile cloud computing and designed an auction-based mechanism for effective resource trading \cite{AJin2017}. Considering that users have individual objectives and the network has a system-wide objective, Zheng \textit{et al.} proposed a gradient-based incentive to motivate truthful information reporting in an algorithm based on alternating direction method of multipliers (ADMM) \cite{ZZheng2017}.  A price incentive based on dual variables, referred to as congestion price, was proposed in \cite{HYaiche2000} for finding the optimal bandwidth allocation. Dual pricing (congestion price) has been adopted in many works, and a recent example is a study on social welfare maximization in a mobile crowdsensing scenario \cite{XDuan2017}.
 
An application of interest for the incentive perspective studies is device-to-device (D2D) communications, in which there can be no traditional network center for D2D links and each user has its own interest.~\footnote{This is a typical case but it could depend on the specific approach of implementing D2D communications.} Su \textit{et al.} proposed a contract based mechanism to motivate truthful reporting in an overlay D2D resource allocation scenario \cite{STSu2017}. Zhao and Song proposed incentive mechanisms to motivate base station data offloading through underlay D2D assisted content distribution \cite{YZhao2017}. A randomized reverse auction was proposed to incentivize content distribution via D2D communications and offload traffic from the base station~\cite{WSongGlobecom2016}. Li \textit{et al.} designed a double auction-based mechanism to incentivize cellular users with alternative D2D communication capabilities to switch to D2D \cite{PLi2016}. Wang \textit{et al.} proposed a resource block exchange mechanism in an overlay D2D scenario to reduce interference experienced by D2D users \cite{CWang2016}. Xu \textit{et al.} exploited an iterative combinatorial auction for resource allocation among D2D links with the objective of sum-rate maximization \cite{CXu2013}. Xu \textit{et al.} applied mechanism design in \cite{LXu2016} for coordinating D2D users to participate in traffic offloading from the BS. Kebriaei \textit{et al.} proposed a double auction to allocate bandwidth in a cellular network with a cognitive D2D user \cite{HKebriaei2016}. Hajiesmaili \textit{et al.} designed an auction to achieve load balancing considering battery limits of D2D devices \cite{AHajiesmaili2017}.   

\subsection{Our Contributions}

Despite the increasing number of related works, there is no general framework that can model a NUM problem with local information, to the best of our knowledge. A close example is \cite{ZZheng2017}, in which general forms of objective functions are used for the network and users. However, the model adopts an optimization perspective and does not capture local information. The objective of this work is to model and study the impact of local information and incentive in solution finding. Two problems are considered. The first is a \textit{user-centric problem} (UCP), a NUM problem with local information in which the network aims to maximize the collective benefit of users. 
The second is a \textit{network-centric problem} (NCP), in which the network aims to maximize its own benefit representing the preference of the network while the users prefer a different solution, e.g., the solution to the UCP. 
After investigating the UCP and the NCP in general forms, the proposed model is applied to underlay D2D communications. The contributions of this work include the followings. 

First, we propose a novel model to capture a NUM problem with local information, in which optimization and incentive perspectives are connected. The proposed model characterizes the interest of a user through two mappings, i.e., an objective function and a valuation function. The objective function maps the optimization variable to a metric such as data rate, energy efficiency, etc. It represents the part that is considered from an optimization perspective. The valuation function maps the above metric to its corresponding value to a user, which can be local information.  
It represents the part that should be handled from an incentive perspective. The model also extends classic NUM problems by allowing each individual user to determine and adopt its own objective locally and, if needed, privately.

Second, dual pricing is studied for the UCP and the insights on when and why dual pricing cannot be used to guarantee truthful reporting are obtained. Dual pricing is of interest since it relates to both a pricing mechanism from the incentive perspective and the dual decomposition technique from the optimization perspective. It is observed in \cite{TT2013} and \cite{RZhang2011} that dual pricing cannot guarantee truthful reporting in the corresponding problems. However, the results of this work provide more insights on dual pricing in general resource allocation problems. It is shown that dual pricing guarantees truthful reporting only when the resource is oversupplied or when a user could not afford any resource. 

Third, we design mechanisms to guarantee truthful reporting when the network has a preferred solution different from the one accepted by the users. We show that, unless this solution is also preferred by at least one user, no mechanism can guarantee truthful reporting and specify a condition under which the network-preferred solution can be achieved. A subsidized exchange mechanism is proposed to motivate truthful reporting and implement the network-preferred solution in a two-user case. The mechanism provides nonnegative and fair utility gain for both users and also a nonnegative gain for the network. 
For the general multiuser case, a much more complicated situation in which more forms of untruthful reporting exist, we propose an iterative extended subsidized exchange mechanism to guarantee that truthful reporting is the best choice for any rational user. 

\section{System Model}\label{s:Sys}

Consider a network with $N$ users, where a user can be a network node, a communication link, etc. The set of users is denoted as $\mathcal{U}=\{i\}_{i= 1}^{N}$. A network center coordinates the resource allocation among users, where the resource allocated to user $i$ is represented by $x_i$.

\subsection{Objective, Valuation, and Local Information}

Each user has an individual objective, such as data rate, energy efficiency, etc. It is assumed that users can have different objectives. This reflects the fact that users may be associated with different applications in a network. For instance, users running throughput-sensitive applications can adopt data rate as their performance metric while users running delay-sensitive applications can adopt delay as their performance metric. Allowing different objectives at different users introduces flexibility in characterizing and coordinating multiuser networks. 

Assume that there are $M$ different objectives and the set of objectives is $\mathcal{O} = \{o_l\}_{l= 1}^{M}$. Denote the objective chosen by user $i$ as $o(i)$. The mapping from $x_i$ to the achieved result on $o(i)$ is denoted as $b_i(x_i)$. 
The mapping $b_i$ is determined by standard formulas, e.g., Shannon capacity. %

For users $i$ and $j$ with the same objective $o_l$, it can be argued that they may value $b_i(x_i)$ and $b_j(x_j)$ differently even if $b_i(x_i) = b_j(x_j)$. For example, 
the same energy efficiency of 100 bits/Hz/J can be of different values to users with 8\% and 20\% respective remaining battery power. This difference reflects different application requirements and resource availability status, e.g., remaining battery power, at the users. To characterize user $i$'s valuation of an achieved objective $b_i(x_i)$, the valuation function $v_i(b_i)$ is introduced. Unlike the mapping $b_i$, $v_i$ is determined individually. 

As aforementioned, a user can choose its own objective and have its unique valuation function based on its application requirements and other local factors such as its battery power. Therefore, it is reasonable to consider the valuation function $v_i(b_i)$ and/or the objective $o_l$ chosen by user $i$ as local/private information. In addition, while $b_i(x_i)$ is given by standard formulas such as the Shannon capacity, parameters in $b_i(x_i)$, e.g., the channel gain, can also be local information. While such information is required for solving an optimization problem in the network, it is not available to the network center unless reported/shared by the users. As a result, truthful reporting on such information from all users can be necessary for finding the solution to a considered problem. The network center, which coordinates and manages the network, is assumed to be not selfish and always truthful. 

The following two subsections introduce the two problems considered in this work.



\subsection{The User-Centric Problem (UCP)}
The network coordinates the resource allocation to maximize the sum-valuation in the network. It is represented by the following problem:   
\begin{subequations}\label{e:PScenarioI} 
\begin{align}
& \mathop{\mathbf{max}}\limits_{\{x_i\}_{i= 1}^N } \quad \sum_i v_i(b_i(x_i))\\
& \mathbf{\quad s.t.}\quad\;   \sum_i x_i \leq X^{max} \label{e:PScenarioIc1} \\
& \qquad\qquad   0 \leq x_i \leq x^{max}, \forall i \in \mathcal{U}, \label{e:PScenarioIc2} 
\end{align}
\end{subequations}
where $X^{max}$ and $x^{max}$ are constants. Note that the adoption of the valuation $v_i(b_i)$ includes the weighted-sum optimization as a special case when $v_i(b_i(x_i)) = \alpha_i b_i(x_i), \forall i$. Problem \eqref{e:PScenarioI} also includes the NUM problems with all users using the same objective, such as a sum-rate optimization problem, as special cases.

Unlike classic NUM problems, local information is captured in this model through the mapping $v_i$ and/or the objective $o_l$ chosen by user $i$. 
Evidently, solving problem \eqref{e:PScenarioI}, in  either a centralized or a distributed approach, requires all users to truthfully reveal their local information. To solve \eqref{e:PScenarioI} in a centralized approach, user $i$ needs to report its chosen objective, the mapping $v_i(b_i)$, and other local information (e.g., the parameter set) to the center. To solve \eqref{e:PScenarioI} in a distributed approach, assuming that the problem is convex and dual decomposition is used, user $i$ should report its solution to the problem of maximizing $v_i(b_i(x_i)) - \lambda x_i$ iteratively, where $\lambda$ is the Lagrange multiplier associated with constraint \eqref{e:PScenarioIc1}. 

If any user reports untruthfully, the solution, assuming that it can be found, might not be the solution to the original problem, i.e., problem~\eqref{e:PScenarioI}. For example, if user $l$ reports $\underline{v}_l(b_l)$ instead of $v_l(b_l)$ \footnote{This corresponds to the case of a centralized optimization approach. In a distributed optimization approach, the equivalent untruthful reporting is that user $l$  reports its local solution based on $\underline{v}_l(b_l)$ instead of $v_l(b_l)$ to the center.} while other users report truthfully, the obtained solution is the solution to the following problem:
\begin{subequations}\label{e:PScenarioIotherP} 
	\begin{align}
	& \mathop{\mathbf{max}}\limits_{\{x_i\}_{i= 1}^N } \quad \underline{v}_l(b_l(x_l)) + \sum_{i\neq l}  v_i(b_i(x_i))\\
	& \mathbf{\quad s.t.}\quad\;   \sum_i x_i \leq X^{max}  \\
	& \qquad\qquad   0 \leq x_i \leq x^{max}, \forall i \in \mathcal{U}.
	\end{align}
\end{subequations}

As each user has its own objective and valuation, assuming truthful reporting can be unconvincing. If the solution to \eqref{e:PScenarioIotherP} is better than the solution to \eqref{e:PScenarioI} from the perspective of user $l$, user $l$ can be motivated to report untruthfully. Therefore, a mechanism is required to incentivize users to report truthfully. For the UCP, a common pricing mechanism, i.e., dual pricing \cite{XDuan2017}, \cite{Namerikawa2015}-\cite{NCLuong2017}, will be studied in details. 

\subsection{The Network-Centric Problem (NCP)}

While problem \eqref{e:PScenarioI} is from the perspective of sum-valuation of users, a different solution can be preferred by the network center. Assume the center prefers the solution to the following problem based on its own objective and valuation $v_c(b_c(\mathbf{x}))$:
\begin{subequations}\label{e:PScenarioII} 
	\begin{align}
	& \mathop{\mathbf{max}}\limits_{\mathbf{x}} \quad  v_c(b_c(\mathbf{x}))\\
	& \mathbf{\quad s.t.}\quad   \sum_i x_i \leq X^{max} \label{e:PScenarioIIc1} \\
	& \qquad\quad\;\;   0 \leq x_i \leq x^{max}, \forall i \in \mathcal{U},
	\end{align}
\end{subequations}
where $\mathbf{x} = [x_1, ..., x_N]$. Denote the solution to \eqref{e:PScenarioII} as $\{x_i^\dagger\}$ and the solution to \eqref{e:PScenarioI} (or any other solution accepted by all users) as $\{x_i^\star\}$. While $\{x_i^\dagger\}$ could be found without help from the users, convincing the users to adopt $\{x_i^\dagger\}$ instead of $\{x_i^\star\}$ requires incentive. \footnote{This is because $\{x_i^\dagger\}$ is a suboptimal solution to \eqref{e:PScenarioI} and generally leads to smaller valuations to a nonempty set of users.} Providing incentive to user $i$ for switching from $x_i^\star$ to $x_i^\dagger$ requires user $i$ to reveal local information regarding $v_i(b_i(x_i^\star))$ and $v_i(b_i(x_i^\dagger))$. A proper mechanism is necessary for the users to report truthfully. For the NCP,  mechanisms are designed to provide incentives for the users to guarantee truthful reporting while switching from $\{x_i^\star\}$ to $\{x_i^\dagger\}$.

In both the UCP and the NCP, the valuation function $v_i(b_i), \forall i$ are assumed to be twice differentiable, strictly concave, nondecreasing, and satisfy $v_i(0) = 0$.\footnote{Most results in this work still apply when the utilities are concave but not strictly concave.} Such valuation models the case of decreasing marginal valuation and is widely used in both classic mechanism design and its applications in the networking field \cite{NNisan2007}. Examples of such utilities can be found in \cite{CJiang2017}. The objective function $b_i(x_i)$ is twice differentiable and can be either concave or unimodal and concave in $[0, x_i^*]$, where $x_i^*$ maximizes $b_i(x_i)$. Section~\ref{s:D2D} will provide detailed examples.

In the remainder of this paper, the focus will be on mechanism analysis and mechanism design based on the following two assumptions: i). All users are selfish but not malicious, i.e., they avoid causing an algorithm to diverge; and ii). A user avoids being identified as untruthful, which can happen if its reports contradict the properties of $v_i(b_i)$ mentioned above.

\vspace{2mm}
\section{Mechanism and Incentive Compatibility}


The concept of mechanism and incentive compatibility is briefly introduced in this section,\footnote{For more details, one can refer to \cite{TBorgers2015}. } while strategies and incentives in the considered system model will be described.

A \textit{mechanism} is a tuple $\{S_1, \dots, S_N, \Gamma(\cdot)\}$ that consists of a set of possible actions $S_i$ for each user and specifies an outcome based on a social choice function $\Gamma(\cdot)$ for each possible combination of actions of users \cite{YN2009}.\footnote{An indirect mechanism is assumed here.} Specifically, in a quasilinear mechanism, $\Gamma(\cdot)$ specifies a transfer $t_i$ (a payment if $t_i >0$ or charge if $t_i < 0$) to user $i$, which renders user $i$'s final utility $u_i$ to be $u_i = v_i + t_i$ (i.e., its valuation of the outcome plus the transfer). 

User $i$ chooses its action from $S_i$ based on its local information and the knowledge on how the outcome and transfer are decided. A common action is to report a value (e.g., a solution, a demand, a quote, etc.). A mechanism is \textit{dominant strategy incentive compatible (DSIC)} if truthful reporting is the best strategy for each user regardless of strategies used by other users \cite{YN2009}. A mechanism is \textit{ex post incentive compatible (EPIC)} if truthful reporting is the best strategy for any user given that other users also report truthfully. 

Several metrics are used to evaluate a mechanism. A mechanism is \textit{efficient} if the outcome maximizes the sum valuation of all users. A mechanism is \textit{individually rational} if each user is guaranteed to have a nonnegative utility. A quasi-linear mechanism is considered \textit{budget-balanced} if $\sum_i t_i \leq 0$. 

For the UCP, reporting occurs in each iteration of distributed optimization while solving problem \eqref{e:PScenarioI}. Specifically 
\begin{itemize}
	\item Iteration $j = 0$. Network (center) chooses an initial value $\lambda^0$ for the Lagrange multiplier $\lambda$ and broadcast $\lambda^0$.
	\item User $i ,\forall i$: expected to report the solution to $\max \big(v_i(x_i) - \lambda^j x_i\big)$ subject to \eqref{e:PScenarioIc2}. 
	\item Network (center) sets $\lambda^{j+1} = \lambda^j + \delta (\sum\limits_i x_i - X^{\max} )$ and broadcast $\lambda^{j+1}$.
	\item Repeat the above two steps until $x_i^j, \forall i$ converge.
\end{itemize}
where $v_i(b_i(x_i))$ is denoted as $v_i(x_i)$ for brevity.

The sequence of reports $\{x_i^j\}, \forall i$ determines whether the algorithm can converge \footnote{Here the term ``converge" refers to ``not diverge" but not necessarily ``converge to the optimal solution".} and whether the solution found is the optimal solution to the original problem \eqref{e:PScenarioI}. The strategy of a user is to choose a report $\underline{x}_i^j$ given the true $x_i^j$ and $\lambda^j$ in each iteration $j$. An example mechanism that specifies the resource allocation outcome and the transfer is given below 
\begin{align}\label{e:MechanismEg}
\left\{
\begin{array}{ll} 
x_i = \tilde{\underline{x}}_i, t_i= f_t(\tilde{\underline{x}}_i) \quad \text{if}\; \{\underline{x}_i^j\}\, \text{converge to}\, \tilde{\underline{x}}_i,{\forall i} \\
x_i = 0,  t_i =0 \qquad\quad\;\, \text{otherwise}
\end{array}
\right.
\end{align}
where $f_t(x)$ is a function for determining transfers. 

 \begin{figure}
	\begin{center}
		\includegraphics[scale = 0.5, angle = 0]{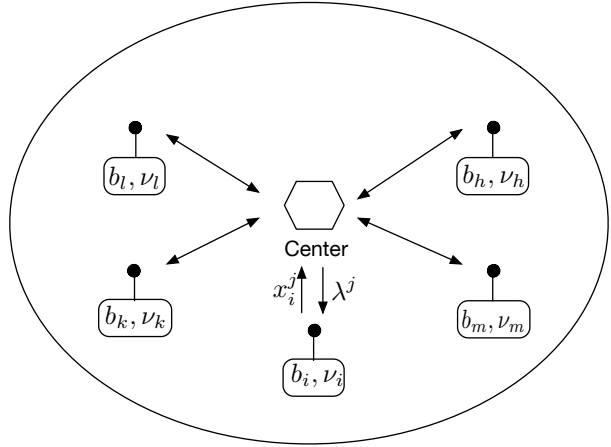}
		\caption{Information Exchange in a distributed solution finding}
		\label{f:sys}
	\end{center}
	\vspace{-4mm}
\end{figure}

A common pricing mechanism for distributed solution finding in the literature is \textit{dual pricing} \cite{XDuan2017}, \cite{Namerikawa2015} - \cite{NCLuong2017}, which interprets and applies the dual variable $\lambda$ in the aforementioned distributed optimization procedure as a per-unit price. Equivalently, $t_i= f_t(\tilde{\underline{x}}_i)  = -\lambda \tilde{\underline{x}}_i$ in \eqref{e:MechanismEg}. It can be seen that dual pricing is individually rational and budget-balanced. However, it is shown by a two-user example in \cite{TT2013} that dual pricing is not incentive compatible when one user is a leader and another is a follower. It is also shown in \cite{RZhang2011} that a dual pricing based auction is not incentive compatible for improving secrecy capacity. The next section will study the dual pricing mechanism and the incentive of users under dual pricing in the considered general scenario.

It should be noted that strategies of a user can be extremely rich, i.e., represented by all mapping from $\{\{x_i^k\}_{k=0}^{j}, \{\lambda^k\}_{k=0}^{j}\}$ to $\underline{x}_i^j$ in iteration $j$. Even if a user makes a decision in each iteration without using historical information, its strategy space still consists of all mappings from $\{x_i^j, \lambda^j\}$ to $\underline{x}_i^j$ in each iteration. However, most of the strategies would trivially lead an optimization algorithm to diverge. Thus, we limit our consideration to the case that users report based on a valuation function $\underline{v}_i(b_i)$ that possesses the same properties as $v_i(b_i)$: strictly concave, twice differentiable, nondecreasing, and $\underline{v}_i(0)=0$. The reporting is truthful if $\underline{v}_i = v_i$. Such reporting strategy is consistent over iterations, allow a distributed algorithm to converge, and is impossible to be identified as untruthful.

\vspace{2mm}
 
\section{Incentive for Truthful Reporting in the UCP: A Study on Dual Pricing}\label{s:Price}

In this section, dual pricing, as a pricing mechanism for solution finding in distributed optimization, is studied. The focus is on whether a user would report truthfully, i.e., $\underline{v}_i =v_i$, so that the solution to the original problem \eqref{e:PScenarioI} can be found. Notations used in this section are given in Table~\ref{t:NotationUCP}. 

Based on the assumptions on $v_i(b_i)$ and $b_i(x_i)$ in Section~\ref{s:Sys}, $v_i(x_i)$ is twice differentiable in $[0, x^{\max}]$ and either concave in $[0, x^{\max}]$ or unimodal and concave in $[0, x_i^*]$, where $x_i^*$ maximizes $v_i(x_i)$, for any $i$ and any objective function $b_i(x_i)$. 

\begin{table}[t!]
	\begin{center}
		\caption{Table of Notations for the UCP}\label{t:NotationUCP}
		{\setlength{\extrarowheight}{1.5pt}
			\begin{tabularx}{0.87\textwidth}{c|c}\hline\hline
				$\mathrm{s}_i$ & reporting strategy of user $i$  \\ \hline
				$\mathrm{s}_{-i}$ & combination of strategies of all users except user $i$ \\ \hline
				$\mathrm{s}_i = \mathrm{s}_i^\mathrm{TR}$ & user $i$ report truthfully, i.e., $\underline{v}_i =v_i$  \\ \hline
				$\mathrm{s}_{-i} = \mathrm{s}_{-i}^\mathrm{TR}$ & other users report truthfully, i.e., $\underline{v}_j =v_j, \forall j\neq i$ \\ \hline
				$\underline{x}_i^\star(\{\mathrm{s}_i, \mathrm{s}_{-i}\})$ & resource allocated to user $i$ given $\{\mathrm{s}_i, \mathrm{s}_{-i}\}$  \\ \hline
				$\underline{u}_i(\mathrm{s}_i, \mathrm{s}_{-i})$ & utility of user $i$ given $\{\mathrm{s}_i, \mathrm{s}_{-i}\}$ \\ \hline
				$\underline{\lambda}^\star(\mathrm{s}_i, \mathrm{s}_{-i})$ & dual price given $\{\mathrm{s}_i, \mathrm{s}_{-i}\}$\\ \hline
				$x_i^\star(\{\mathrm{s}_i^\mathrm{TR}, \mathrm{s}_{-i}^\mathrm{TR}\})$  & 	$\underline{x}_i^\star(\{\mathrm{s}_i, \mathrm{s}_{-i}\})$ when $\mathrm{s}_i = \mathrm{s}_i^\mathrm{TR}$ and $\mathrm{s}_{-i} = \mathrm{s}_{-i}^\mathrm{TR}$\\ \hline
				$u_i(\mathrm{s}_i^\mathrm{TR}, \mathrm{s}_{-i}^\mathrm{TR})$  & $\underline{u}_i(\mathrm{s}_i, \mathrm{s}_{-i})$ when $\mathrm{s}_i = \mathrm{s}_i^\mathrm{TR}$ and $\mathrm{s}_{-i} = \mathrm{s}_{-i}^\mathrm{TR}$\\ \hline
				$\lambda^\star(\mathrm{s}_i^\mathrm{TR}, \mathrm{s}_{-i}^\mathrm{TR})$ & 	$\underline{\lambda}^\star(\mathrm{s}_i, \mathrm{s}_{-i})$ when $\mathrm{s}_i = \mathrm{s}_i^\mathrm{TR}$ and $\mathrm{s}_{-i} = \mathrm{s}_{-i}^\mathrm{TR}$\\ 
				\hline\hline
			\end{tabularx}
		}
	\end{center}
	\vspace{-0.2cm}
\end{table}

\textbf{Lemma~1}: The utility of user $i$ satisfies $\underline{u}_i(\{\mathrm{s}_i, \mathrm{s}_{-i}^\mathrm{TR}\}) \leq u_i(\{\mathrm{s}_i^\mathrm{TR}, \mathrm{s}_{-i}^\mathrm{TR}\})$ for any reporting strategy $\mathrm{s}_i$ such that $\underline{\lambda}^\star(\{\mathrm{s}_i, \mathrm{s}_{-i}^\mathrm{TR}\}) > \lambda^\star(\{\mathrm{s}_i^\mathrm{TR}, \mathrm{s}_{-i}^\mathrm{TR}\})$,  where the strict inequality holds when $\underline{x}_i^\star(\{\mathrm{s}_i, \mathrm{s}_{-i}^\mathrm{TR}\})> 0$.  

\textit{Proof}: Please refer to Section~\ref{p:Lemma1} in Appendix.

Lemma 1 shows that, when a user intends to report untruthfully, it should avoid any untruthful reporting strategy that can unilaterally drive $\lambda$ up even if it can acquire more resource by using such a strategy. 


\textbf{Lemma~2}: User $i$ can unilaterally affect the price through untruthful reporting such that $\underline{\lambda}^\star (\mathrm{s}_i,  \mathrm{s}_{-i}^\mathrm{TR}) < \lambda^\star(\mathrm{s}_i^\mathrm{TR}, \mathrm{s}_{-i}^\mathrm{TR})$ with $\mathrm{s}_i \neq \mathrm{s}_i^\mathrm{TR}$ as long as $x_i^\star(\mathrm{s}_i^\mathrm{TR}, \mathrm{s}_{-i}^\mathrm{TR}) \neq 0$.

\textit{Proof}: Please refer to Section~\ref{p:Lemma2} in Appendix.

Lemma~2 shows that each user has the ability to unilaterally drive $\lambda$ down through untruthful reporting as long as its share of the resource when everyone reports truthfully, i.e., $x_i^\star(\mathrm{s}_i^\mathrm{TR}, \mathrm{s}_{-i}^\mathrm{TR})$, is nonzero. 

\textbf{Lemma 3}: Unless $\sum_i x_i^\star(\mathrm{s}_i^\mathrm{TR}, \mathrm{s}_{-i}^\mathrm{TR}) < X^{\max}$, truthful reporting is not optimal for any user $i$ with $x_i^\star(\mathrm{s}_i^\mathrm{TR}, \mathrm{s}_{-i}^\mathrm{TR}) \neq 0$, for any given combination of valuation functions $\{v_i(x_i)\}_{\forall i}$, assuming others report truthfully. A $\underline{x}_i^\star \in (0,  x_i^\star)$ and a corresponding $\underline{\lambda}^\star  \in (0, \lambda^\star)$ always exist such that 
\begin{align}
\underline{u}_i(\mathrm{s}_i,\! \mathrm{s}_{-i}^\mathrm{TR}) \!=\! v_i(\underline{x}_i^\star) \!-\! \underline{\lambda}^\star \underline{x}_i^\star \!>\! v_i(x_i^\star) \!- \! \lambda^\star x_i^\star \!=\! u_i(\mathrm{s}_i^\mathrm{TR}\!,\! \mathrm{s}_{-i}^\mathrm{TR}),
\end{align}	
where $\underline{x}_i^\star$,  $x_i^\star$,  $\underline{\lambda}^\star$, and  $\lambda^\star$ are abbreviations of $\underline{x}_i^\star(\{\mathrm{s}_i, \mathrm{s}_{-i}^\mathrm{TR}\})$, $x_i^\star(\{\mathrm{s}_i^\mathrm{TR}, \mathrm{s}_{-i}^\mathrm{TR}\})$, $\underline{\lambda}^\star(\mathrm{s}_i, \mathrm{s}_{-i}^\mathrm{TR})$, and $\lambda^\star(\mathrm{s}_i^\mathrm{TR}, \mathrm{s}_{-i}^\mathrm{TR})$, respectively. 

\textit{Proof}: Please refer to Section~\ref{p:Lemma3} in Appendix.

\begin{figure}[t!]
	\begin{center}
		\includegraphics[scale = 0.58, angle = 0]{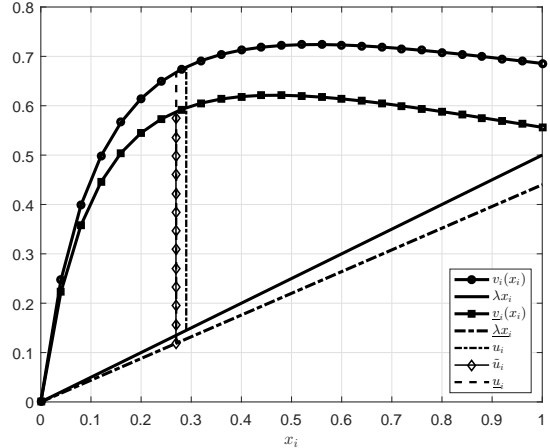}
		\caption{An illustration on how a user could benefit from reporting untruthfully. The variable $x_i$ represents the resource allocated to user $i$. Given any $x_i$, the corresponding $y$ coordinates on the two curves represent the truthful and untruthful valuations, respectively. The corresponding $y$ coordinates on the two oblique lines stand for the costs under truthful and untruthful reporting, respectively. The gaps represent utilities of user $i$ (please refer to the paragraph following Lemma~3).}
		\label{f:PriceCh}
	\end{center}
	\vspace{-4mm}
\end{figure}

Lemma~3 shows that an untruthful reporting strategy can result in a smaller share of resource but yield a larger utility for a user due to a lower price. Such a strategy is related to \textit{demand reduction}, and discussions on a simpler discrete case in which resources are multiple indivisible items for sale can be found in \cite{TR2014Lec}. An illustration of Lemma~3 is shown in Fig.~\ref{f:PriceCh}, where $v_i$ and $\lambda_i x_i$ are the true valuation function and the cost with truthful reporting, respectively, and $\underline{v}_i$ and $\underline{\lambda}_i \underline{x}_i$ are an untruthfully-reported valuation and the corresponding cost, respectively. The variables $u_i$, $\tilde{u}_i$, and $\underline{u}_i$ are the utility of user $i$ with truthful reporting, the utility of user $i$ if $\underline{v}_i$ were the truthful valuation, and the utility actually acquired by user $i$ with untruthful reporting based on $\underline{v}_i$, respectively.

Lemma~1 to Lemma~3 provide insights into dual pricing as a mechanism for a distributed resource allocation problem step by step. First, it is shown by Lemma~1 that dual pricing can prevent untruthful reporting that would intensify the competition and drive the price $\lambda$ up. Next, Lemma~2 shows that users that would be allocated a nonzero share of resource when everyone report truthfully can unilaterally drive $\lambda$ downward. Lemma~3 shows that such a user can benefit from driving $\lambda$ downward when all other users report truthfully except for trivial cases in which the resource is oversupplied, i.e., $\sum_i x_i^\star(\mathrm{s}_i^\mathrm{TR}, \mathrm{s}_{-i}^\mathrm{TR}) < X^{\max}$. The condition that $x_i^\star \neq 0$ in Lemmas~2~and~3 implies that truthful reporting is optimal for a user when it cannot afford any resource under dual pricing. The above insights lead to the following remark.

\textit{Remark 1}: Under dual pricing, reporting truthfully is optimal for a user only when the resource is oversupplied or when it cannot afford any resource.  

\textbf{Theorem~1}: Dual pricing, when used as a pricing mechanism, is not EPIC.

\textit{Proof}: Combining Lemma~1 to Lemma~3 proves this result. 

The results in Lemma~1 to Lemma~3 are obtained without needing to know the specific local valuations of the users. It implies that a rational user can also obtain the above insights while choosing its reporting strategy and realize that truthful reporting is unlikely to be optimal.

As dual pricing is not incentive compatible, a different mechanism is needed to guarantee truthful reporting in a distributed optimization. The extension of classic Groves mechanisms to algorithmic mechanism design can be adopted here\cite{TT2013}. The mechanism is implemented through an algorithm with dual-decomposition based iterations while, in addition to their local solutions,  users are also required to report their valuations in each iteration. The network center updates the dual variable, performs the convergence check, and charges each user a VCG based cost instead of a dual-price based cost at convergence. For any given iteration, the mechanism is not incentive compatible. However, the mechanism is shown to be \textit{asymptotically DSIC}, which yields a diminishing return for an untruthful reporting user as the algorithm proceeds. 
We remark that the same extension of Groves mechanisms is also asymptotically DSIC when a projected gradient ascent is used, in which case the coordinator is not computationally involved but only responsible for checking convergence and executing the final allocation solution. With dual pricing, however, the result would be different.

\section{Incentive for Truthful Reporting in the NCP: Condition and Mechanisms} \label{s:ImplementMec}

The preceding section investigates incentive for truthful reporting while \textit{solving} the UCP. This section investigates incentive for truthful reporting while \textit{implementing} the solution of the NCP. Recall that the center prefers a different solution, i.e., $\{x_i^{\dagger}\}$, from the one accepted by the users, i.e., $\{x_i^{\star}\}$ in the NCP. The solution $\{x_i^{\dagger}\}$ is the solution to problem \eqref{e:PScenarioII}. Although the center can solve problem \eqref{e:PScenarioII} without help from the users, incentives need to be provided to the users for a consensus to adopt $\{x_i^{\dagger}\}$ instead of  $\{x_i^{\star}\}$. To determine the incentives, reports from the users regarding their valuations are necessary, and therefore truthful reporting is a concern. 

Denote $\nu(\mathbf{x}) = v_c(b_c(\mathbf{x}))$. The study is based on the assumption that, while $\mathbf{x}^\dagger$ is optimal to the center, the center is also willing to adopt any intermediate solution $\mathbf{x}$ at the cost of providing transfers $\{t_i\}$ to the users as long as $\nu(\mathbf{x}) -  \sum_i t_i(x_i) \geq \nu(\mathbf{x}^\star)$. A mechanism can \textit{implement} $\mathbf{x}^\dagger$ if users prefer $\mathbf{x}^\dagger$ after receiving the transfers, i.e., $v_i(x_i^{\dagger}) + t_i\geq v_i(x_i^{\star}), \forall i$. Properties such as convexity are not assumed for $\nu(\mathbf{x})$ as long as the center can solve problem \eqref{e:PScenarioII}.


With the above assumptions, whether $\{x^\dagger\}$ can be implemented or not and what mechanism can be used to implement $\{x^\dagger\}$ are investigated.

%

\textbf{Lemma 4}: If $x_i^\dagger  < x_i^\star, \forall i$, no mechanism can be both incentive compatible and able to specify a condition under which the implementation of $\mathbf{x}^\dagger$ is guaranteed. 

{\it Proof}: Please refer to Section~\ref{p:Lemma4} in Appendix. 

Lemma 4 shows that at least one user, in addition to the center, should prefer $x_i^\dagger$ to $x_i^\star$. Otherwise, no mechanism can be incentive compatible while guaranteeing the implementation of $\mathbf{x}^\dagger$.

Next consider the case that $x_i^\dagger  < x_i^\star$ for a  set of $i$ denoted as $\mathcal{S}_1$ and $x_i^\dagger  > x_i^\star$ for a nonempty set of $i$ denoted as $\mathcal{S}_2$. 

Consider a two-user example first. Assume $v_1(x_1^\star) > v_1(x_1^\dagger)$ and $v_2(x_2^\star) < v_2(x_2^\dagger)$. 
A subsidized exchange mechanism (SEM) for motivating users to switch to $\{x_i^\dagger\}$ from $\{x_i^\star\}$ is proposed in Algorithm~\ref{a:SEM}. In this algorithm, Line 3 and Line 4 are executed by the users and the rest by the center.

 \begin{algorithm}[ht!]
	\caption{Subsidized Exchange Mechanism (SEM)}\label{a:SEM}
	\begin{algorithmic}[1]
		\renewcommand{\algorithmicrequire}{\textbf{Input:}} 
		\renewcommand{\algorithmicensure}{\textbf{Output:}}
		\REQUIRE  $\mathbf{x}^\dagger, \mathbf{x}^\star$
		\ENSURE  $\mathbf{x}$
		\STATE Calculate $s_c =  \nu(\mathbf{x}^\dagger) - \nu(\mathbf{x}^\star) $ and determine a threshold $\alpha \in (0, 1/2]$
		\STATE Announce $x_1^\dagger$ and $x_2^\dagger$ to users 1 and 2, respectively.
		\STATE User 1 reports $\rho$, i.e., the compensation it requires to implement $x_1^\dagger$, to the center.
		\STATE User 2 reports $\varphi$, i.e., the price it is willing to pay to implement $x_2^\dagger$, to the center.
		\STATE Compare $\varphi + \alpha s_c$ with $\rho$. 
		\IF {$\varphi + \alpha s_c  < \rho$}
		\STATE abort with $\mathbf{x} = \mathbf{x}^\star$
		\ELSE 
		\STATE Charge user 2 the amount of $\rho - \alpha s_c $ and pay user 1 the amount of $\varphi + \alpha s_c$
		\STATE $\mathbf{x} = \mathbf{x}^\dagger$
		\ENDIF
		\RETURN $\mathbf{x}$
	\end{algorithmic}
\end{algorithm}		


\textbf{Lemma 5}: The mechanism in Algorithm~\ref{a:SEM} can implement $\{x_1^\dagger, x_2^\dagger\}$ under the condition $\alpha s_c \geq \rho - \varphi$. Moreover, the proposed mechanism is DSIC and individually rational. 

{\it Proof}: Please refer to Section~\ref{p:Lemma5} in Appendix.

Since $\{x_i^\star\}$ maximizes $\sum_i v_i(x_i)$, it can be seen that  $\rho^\mathrm{TR} = v_1(x_1^\star) -  v_1(x_1^\dagger) >  v_2(x_2^\dagger) - v_2(x_2^\star) = \varphi^\mathrm{TR}$ since $\sum_i v_i$ is strictly concave. Consequently, $\varphi + \alpha s_c \geq \rho$ is possible only if $\alpha s_c > 0$ given that $\rho \geq \rho^\mathrm{TR}$ and $\varphi \leq \varphi^\mathrm{TR}$ for any rational users. Therefore, the center must subsidize an exchange for it to be successful, i.e.,  $\varphi + \alpha s_c  \geq \rho$. In addition to being DSIC and individually rational, the SEM also gives the center a nonnegative gain after deducting the subsidy. Specifically, the gain of the center is  
\begin{align}
\pi_c &= s_c - \big(\varphi + \alpha s_c - (\rho - \alpha s_c)\big) \nonumber \\
&= (1 - 2 \alpha)s_c + \rho - \varphi > (1 - 2 \alpha) s_c
\end{align}
where $s_c =  \nu(\mathbf{x}^\dagger) - \nu(\mathbf{x}^\star)$ represents the value of a successful exchange to the center. The terms $\varphi + \alpha s_c$ and $\rho - \alpha s_c$ are the amounts that the center pays and receives in SEM, respectively, with the difference being the subsidy. The inequality follows from the fact that $\rho \geq \rho^\mathrm{TR} \geq \varphi^\mathrm{TR} \geq \varphi$. The choice of $\alpha$ in $(0, 1/2]$ in Algorithm~\ref{a:SEM} guarantees that $\pi_c > 0$ in an exchange. 

Insights can be drawn from comparing the cases with and without considering information availability and incentive. If $\rho^\mathrm{TR}$ and $\varphi^\mathrm{TR}$ were information known to the center, the center could coordinate a successful exchange as long as $s_c \geq \rho^\mathrm{TR} - \varphi^\mathrm{TR}$.~\footnote{Specifically, the center could offer $\rho^\mathrm{TR} + \epsilon$ to user 1 and charge $\varphi^\mathrm{TR} - \epsilon$ from user 2  with an nonnegative $\epsilon$ that can be arbitrarily small or even 0.} By contrast, with $\rho^\mathrm{TR}$ and $\varphi^\mathrm{TR}$ being local information, a successful exchange occurs when $\alpha s_c \geq \rho^\mathrm{TR} - \varphi^\mathrm{TR}$ in SEM. Since $\alpha \leq 1/2$, the condition that $\alpha s_c \geq \rho^\mathrm{TR} - \varphi^\mathrm{TR}$ is more stringent than the condition $s_c \geq \rho^\mathrm{TR} - \varphi^\mathrm{TR}$. Consequently, the chance of implementing $\{x_1^\dagger, x_2^\dagger\}$ becomes smaller. It shows that, when $\rho^\mathrm{TR}$ and $\varphi^\mathrm{TR}$ are local information, the center needs to provide more subsidy (compared to $\rho^\mathrm{TR} - \varphi^\mathrm{TR}$) to incentivize an exchange. 
 Thus, the more stringent requirement on $s_c$ and, as a result, a smaller chance of implementing $\{x_i^\dagger\}$ represent the cost of dealing with local information.


It should be noted that the parameter $\alpha$ must be determined before the center receives $\varphi$ and $\rho$. The choice of $\alpha$ corresponds to a trade-off for the center. A smaller $\alpha$ leads to a smaller chance of success in implementing $\{x_1^\dagger, x_2^\dagger\}$ but a larger $\pi_c/s_c$. By contrast, a larger $\alpha$ increases the chance of success yet leads to a smaller $\pi_c/s_c$ for the center. 

{\it Remark 2}: The SEM provides an equal and fair utility gain, i.e., $\varphi^\mathrm{TR} - \rho^\mathrm{TR} + \alpha s_c$, to both users in any successful exchange.

{\it Remark 3}: The benchmark allocation $\{x_i^{\star}\}$ can be either the solution to problem \eqref{e:PScenarioI} or any other solution accepted by both users as long as at least one user prefers $\{x_i^{\dagger}\}$ to $\{x_i^{\star}\}$ (otherwise Lemma~4 applies).

The next step is to consider a multiuser case. Generally, it is impossible to determine whether $\{x_i^{\dagger}\}$ can be implemented or not in one shot due to the coupling between $\nu(\mathbf{x})$ and $v_i(x_i), \forall i$. Therefore, an iterative algorithm is proposed, as an extension of SEM, in which the center coordinates the users to gradually migrate from $\{x_i^{\star}\}$ toward $\{x_i^{\dagger}\}$ through a sequence of subsidized exchanges. The proposed extended subsidized exchange mechanism (ESEM) is given in Algorithm~\ref{a:ESEM}. The algorithm ends up with either the implementation of $\{x_i^{\dagger}\}$ or a solution no worse than $\{x_i^{\star}\}$ to all users as well as the center (both with the subsidies taken into account). 

\begin{figure}
	\begin{center}
		\includegraphics[scale = 0.48, angle = 0]{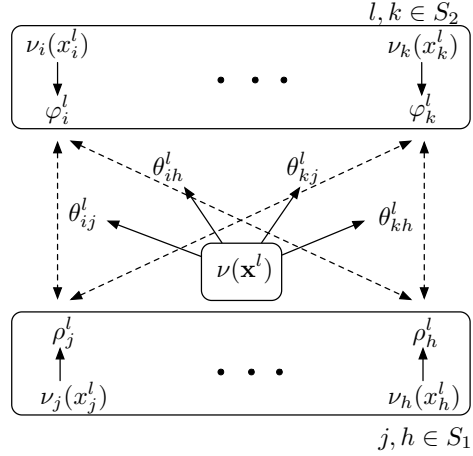}
		\caption{Illustration of variable relations in iteration $l$ (ESEM).}
		\label{f:ESEM}
	\end{center}
	\vspace{-4mm}
\end{figure}

Let $l$ represent the iteration index, and $x_i^l$ represent $x_i$ in iteration $l$. Initialize $x_i^0 = x_i^{\star}, \forall i$.  Define two sets  $\mathrm{S}_1^l= \{i|x_i^{\dagger} < x_i^l, i\in \mathcal{U}\}$ and  $\mathrm{S}_2^l = \{i| x_i^{\dagger} > x_i^l, i\in \mathcal{U}\}$. In the $l$th iteration, user $i, \forall i \in \mathrm{S}_1^l$ reports $\rho_i^l$, its loss from implementing $x_i^l - \delta_l$ compared to implementing $x_i^l$, and user $j, \forall j \in \mathrm{S}_2^l$ reports $\varphi_j^l$, its gain from implementing $x_j^l + \delta_l$ compared to implementing $x_j^l$. For each combination of $i \in \mathrm{S}_1^l$ and $j \in \mathrm{S}_2^l$, the center calculates its valuation of the tentative exchange between users $i$ and $j$ as 
\begin{align}\label{e:theta}
\theta_{ij}^l = \nu(\mathbf{\bar{x}}^{l+1, (ij)}) - \nu(\mathbf{x}^{l}), 
\end{align}
where $\mathbf{\bar{x}}^{l+1,(ij)} = [\bar{x}_1^{l+1,(ij)}, \dots, \bar{x}_N^{l+1,(ij)}]$ with 
\begin{align}
\bar{x}_k^{l+1,(ij)} =\left\{
\begin{array}{ll}
x_k^{l} -  \delta_l \qquad\qquad\qquad\qquad \text{if}\; k = i \\
x_k^{l} +  \delta_l \qquad\qquad\qquad\qquad \text{if}\; k = j \\
x_k^{l} \qquad\qquad\qquad\qquad\quad\; \text{otherwise}.\\
\end{array}
\right.
\end{align}

Define $\Theta^l = \{\theta_{ij}^l\}$. Define 
\begin{align}\label{e:psi}
\psi_{ij}^l = \alpha^l \theta_{ij}^l + \varphi_j^l - \rho_i^l
\end{align}
and $\varPsi^l = \{\psi_{ij}^l\}$. For any given $i \in \mathrm{S}_1^l$ and $j \in \mathrm{S}_2^l$ in iteration $l$, define the following sets 
\begin{subequations}\label{e:OmegaSets}
\begin{align}
\Omega_i^l = \{j|j \in \mathrm{S}_2^l, \psi_{ij}^l >0 \} \\
\Omega_j^l = \{i|i \in \mathrm{S}_1^l, \psi_{ij}^l >0 \},
\end{align}
\end{subequations}
For simplicity, define a $|\mathrm{S}_1^l|\times |\mathrm{S}_2^l|$ matrix $\mathrm{W}^l$ with its elements determined by
\begin{align}\label{e:Wmatrix}
\mathrm{W}_{i,j}^l &=\left\{
\begin{array}{ll} 
1, \quad\quad\quad\quad\quad\quad \text{if}\; j \in \Omega_i^l \\
0, \quad\quad\quad\quad\quad\quad  \text{otherwise.}\;\\
\end{array}
\right.
\end{align}
Define
\begin{align}\label{e: RiPlus}
\mathrm{R}_{i+}^l = \{i|i \in \mathrm{S}_1^l, \sum\limits_j \mathrm{W}_{i,j}^l \neq 0\} 
\end{align}
as the set of row indices corresponding to the rows of $\mathrm{W}_{i,j}^l $ with at least one nonzero element. With the above definitions, the ESEM is given in Algorithm~\ref{a:ESEM}, in which Line~4 and Line~7 are the steps executed by the users and others  by the center.

\textbf{Theorem 2:} The ESEM in Algorithm~\ref{a:ESEM} is EPIC and individually rational if $\{\alpha^l\}_{\forall l}$ is chosen such that $\{\alpha^l \theta_{\tilde{i}\tilde{j}}^l\}_{\forall l}$ is a nonincreasing sequence.

\textit{Proof}: See Section~\ref{p:Theorem2} in Appendix. 

It is worth noting that there exist more forms of untruthful reporting in an iterative procedure than in a one-shot exchange. Proved to be incentive compatible, ESEM can prevent all of the following cases of untruthful report including: (i) untruthfully report a larger $\rho_i^l$ or $\phi_i^l$ in iteration $l$ to achieve larger benefit, (ii) untruthfully report a smaller $\rho_i^l$ or $\phi_i^l$ in iteration $l$ to achieve larger benefit, and (iii) skipping an iteration to achieve a larger benefit when $\delta_l$ changes. 

 \begin{algorithm}[H]
	\caption{Extended Subsidized Exchange Mechanism (ESEM)}\label{a:ESEM}
	\begin{algorithmic}[1]
		\renewcommand{\algorithmicrequire}{\textbf{Input:}} 
		\renewcommand{\algorithmicensure}{\textbf{Output:}}
		\REQUIRE  $\mathrm{S}_1^0$, $\mathrm{S}_2^0, \mathbf{x}^\dagger, \mathbf{x}^\star$
		\ENSURE  $\mathbf{x}^l$
		\\ \textit{Initialization}: Set $l = 0$, $x_i^0 = x_i^{\star},\forall i$. Set $\delta^0$.
		\WHILE{(true)}
		\STATE Determine $\alpha^l$ and announces $\delta^l$ to users
		\FOR {User $i = 1$ to $N$}
		\IF {$ i \in\mathrm{S}_1^l$}
		\STATE Report $\rho_i^l = v_i(x_i^l) - v_i(x_i^l - \delta_l)$ to the center
		\ELSIF {$ i \in\mathrm{S}_2^l$}
		\STATE Report $\varphi_j^l = v_j(x_j^l + \delta_l) - v_j(x_j^l)$ to the center
		\ENDIF
		\ENDFOR
		\STATE Calculate $\theta_{ij}^l$ and $\psi_{ij}^l, \forall i \in \mathrm{S}_1^l, \forall j \in \mathrm{S}_2^l$ using \eqref{e:theta} and \eqref{e:psi}. Calculate $\mathrm{W}^l$ using \eqref{e:OmegaSets} and \eqref{e:Wmatrix}. 
		\IF {$\mathrm{W}^l = \mathbf{0}$}
		\STATE exit with $\mathbf{x}^l$
		\ENDIF
		\STATE Set $\mathrm{R}_{i+}^l$ as in \eqref{e: RiPlus} and $\mathrm{R}_{j+}^l = \mathrm{S}_2^l$
		\WHILE {($\mathrm{R}_{i+}^l\neq \varnothing $ and $\mathrm{R}_{j+}^l \neq \varnothing$) }
		\STATE Select an $i$, denoted as $\tilde{i}$, from $\mathrm{R}_{i+}^l$ based on a uniform  distribution
		\STATE Select a $j$, denoted as $\tilde{j}$, from $\mathrm{S}_2^l$ based on a uniform distribution
		\IF { $W_{\tilde{i}, \tilde{j}}^l = 0$ }
		\STATE $\mathrm{R}_{i+}^l = \mathrm{R}_{i+}^l / \{\tilde{i}\}$
		\STATE $\mathrm{R}_{j+}^l = \mathrm{R}_{j+}^l / \{\tilde{j}\}$
		\ELSE 
		\STATE Charge user $\tilde{j}$ the amount $\rho_{\tilde{i}} - \alpha^l \theta_{\tilde{i}\tilde{j}}^l$
		\STATE Compensate user $\tilde{i}$ the amount $\varphi_{\tilde{j}} + \alpha^l  \theta_{\tilde{i}\tilde{j}}^l$
		\STATE Set $\mathbf{x}^{l+1} = \mathbf{x}^l$, $\mathrm{S}_1^{l+1} = \mathrm{S}_1^l$, and $\mathrm{S}_2^{l+1} = \mathrm{S}_2^l$
		\STATE Update $x_{\tilde{i}}^{l+1} = x_{\tilde{i}}^l - \delta_l$ and $x_{\tilde{j}}^{l+1} = x_{\tilde{j}}^l + \delta_l$
		\STATE $\mathrm{S}_1^{l+1} = \mathrm{S}_1^l/\{\tilde{i}\}$ if $x_{\tilde{i}}^{l+1} =  x_{\tilde{i}}^\dagger$
		\STATE $\mathrm{S}_2^{l+1} = \mathrm{S}_2^l/\{\tilde{j}\}$ if $x_{\tilde{j}}^{l+1} =  x_{\tilde{j}}^\dagger$
		\STATE break the inner while loop
		\ENDIF
		\ENDWHILE
		\IF {$\mathrm{R}_{i+}^l = \varnothing $ or $\mathrm{R}_{j+}^l = \varnothing$ or $\mathrm{S}_1^{l+1} = \varnothing$}
		\STATE Set $l = l + 1$ and exit with $\mathbf{x}^{l}$
		\ENDIF
		\STATE Set $l = l + 1$ and update $\delta_{l}$ 
		\ENDWHILE
		\RETURN $\mathbf{x}^{l}$
	\end{algorithmic}
\end{algorithm}

\section{Application on Underlay D2D communications}\label{s:D2D}


The modeling, analysis, and mechanisms from the preceding sections can be applied in networks featuring a distributed/semi-distributed structure in which users have individual utilities and local information. In this section, the UCP and the NCP will be re-modeled in an application of underlay D2D communications. Other applications in which the proposed model and results may apply include but are not limited to: spectrum leasing in cognitive radio, mobile crowdsensing, cooperative data dissemination in vehicular ad-hoc networks, demand response in smart grid, and cooperative caching.   

\begin{figure}
	\begin{center}
		\includegraphics[scale = 0.51,angle = 0]{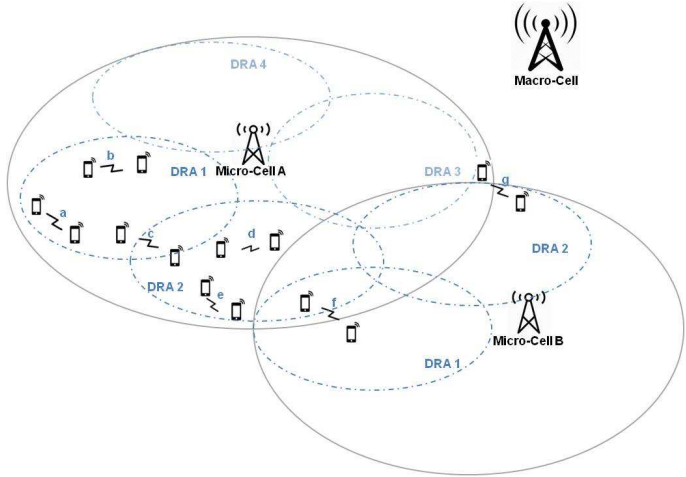}
		\caption{The underlay D2D application scenario. The abbreviation DRA stands for D2D reuse area.}
		\label{f:D2Dsys}
	\end{center}
	\vspace{-4mm}
\end{figure}

Consider a cellular network in an urban environment with dense microcells as shown in Fig.~\ref{f:D2Dsys}. An underlay based D2D communications is adopted, and resource blocks (RBs) for cellular communications are reused by D2D communications.  A microcell is divided into D2D reuse areas, and RBs can be reused by different D2D links in different reuse areas.\footnote{Considering the fact that diameter of a micro-cell is on the kilometer level while the diameter of D2D communications is mostly 25 meters or less \cite{CXu2013}\cite{LSong2014}\cite{ZZhou2014}, allowing only one D2D link to communicate on each RB would lead to a waste of spectral resource.} For cross-area D2D links, e.g., link $c$ in Fig.~\ref{f:D2Dsys}, the D2D link can communicate when the microcell BS (microBS) can assign an RB that is available in both areas. D2D links associated with different microBS cannot communicate directly. To control the interference from D2D to cellular and other D2D communications, two assumptions are made. First, the maximum transmit power allowed for D2D communications, denoted as $p^{\max}$, is smaller than that allowed for cellular communications. Second, the total transmit power of all D2D communications reusing the same RB in each microcell is limited by  $P^{\max}$.        


Denote the $k$th D2D RB and the set of D2D RBs as $b_k$ and $\mathcal{B}$, respectively, and the size of $\mathcal{B}$ as $N_B$. Also denote $\mathcal{L}_m(b_k),  k \in \{1, \dots, N_B\}$ as the set of D2D links allocated with RB $b_k$ in all areas of microcell $m$ and the size of $\mathcal{L}_m(b_k)$ as $L_m^{b_k}$. Let $\mathcal{C}_m$ represent the set of all neighbor microcells of microcell $m$. 
A D2D link in $\mathcal{L}_m(b_k)$ may receive interference from four sources: other D2D links in $\mathcal{L}_m(b_k)$, D2D links in $\mathcal{L}_s(b_k), s\in \mathcal{C}_m$, cellular links using $b_k$ in microcell $m$, and cellular links using $b_k$ in microcells in $\mathcal{C}_m$. 
The received signal of a D2D link in $\mathcal{L}_m(b_k)$ is 
\begin{align}\label{e:RxSignal}
y_{m,i} = & \sqrt{p_{m,i}} h_{m,i}^{b_k}x_{m,i} + \sum\limits_{j\in \mathcal{L}_m(b_k), j\neq i}\sqrt{p_{m,j}} h_{m,j,i}^{b_k}x_{m,j} \nonumber \\ 
& + \sum\limits_{s \in \mathcal{C}_m}\sum\limits_{j \in \mathcal{L}_s(b_k)} \sqrt{p_{s,j}} h_{s,j,i}^{b_k} x_{s,j} + \sqrt{p_{m}^\mathrm{c}} h_{m,i}^{\mathrm{c},b_k}x_{m}^\mathrm{c} \nonumber \\
& + \sum\limits_{s \in \mathcal{C}_m}\sqrt{p_{s}^\mathrm{c}} h_{s,i}^{\mathrm{c},b_k}x_{s}^\mathrm{c} + n_{m,i},
\end{align} 
where $x_{m, i}$, $x_{m, j}$, and $x_{s, j}$ represent the transmitted signal of the $i$th D2D link using $b_k$ in microcell $m$, the $j$th D2D link using $b_k$ in microcell $m$, and the $j$th D2D link using $b_k$ in microcell $s$, respectively. The variables $p_{m, i}$, $p_{m, j}$, and $p_{s, j}$ represent the transmit power of the $i$th D2D link using $b_k$ in microcell $m$, the $j$th D2D link using $b_k$ in microcell $m$, and the $j$th D2D link using $b_k$ in microcell $s$, respectively. The variables $x_{m}^\mathrm{c}$, $x_{s}^\mathrm{c}$, $p_{m}^\mathrm{c}$, and $p_{s}^\mathrm{c}$ denote the transmitted signal of the cellular links using $b_k$ in microcells $m$ and $s$, and the corresponding transmit power of the cellular links, respectively. The variable $h_{m,i}^{b_k}$ represents the channel from the target transmitter to the target receiver of the $i$th D2D link in microcell $m$. The variables $h_{m,j,i}^{b_k}$ and $h_{s,j,i}^{b_k}$ represent the interference channels from the transmitter of the $j$th D2D link using $b_k$ in microcell $m$ and from the transmitter of the $j$th D2D link using $b_k$ in microcell $s$ to the target receiver of the $i$th D2D link in microcell $m$, respectively. The variables $h_{m,i}^{\mathrm{c}, b_k}$ and $h_{s,i}^{\mathrm{c}, b_k}$ are the interference channels from the transmitters of the cellular links using $b_k$ in microcells $m$ and $s$ to the target receiver of the $i$th D2D link in microcell $m$, respectively. The variable $n_{m,i}$ represents the additive Gaussian noise.

The first item in \eqref{e:RxSignal} is the intended signal, and the second to the fifth items represent the interference from other D2D links in the same microcell, D2D links in the neighboring cells, the cellular link in the same microcell, and cellular links in the neighboring microcells, respectively. The channels incorporate Rayleigh fading and large scale propagation path-loss. It is assumed that D2D links have no preference in particular RBs. 

The data rate corresponding to Eq.~\eqref{e:RxSignal} is given as
\begin{align}\label{e:rate}
r_{m,i}(p_{m,i}) = \log \bigg(1 + \frac{g_{m,i}^{b_k} p_{m,i}}{\sigma_0^2 + p_{m,i}^{I}}\bigg),
\end{align}    
where $g_{m,i}^{b_k} = |h_{m,i}^{b_k}|^2$ represents the channel power gain, and 
\begin{align}
p_{m,i}^{I} 
&= \sum\limits_{j\in \mathcal{L}_m(b_k), j\neq i} p_{m,j} g_{m,j,i}^{b_k} + \sum\limits_{s \in \mathcal{C}_m}\sum\limits_{j \in L_s(b_k)} p_{s,j} g_{s,j,i}^{b_k} \nonumber \\ &+ p_{m}^\mathrm{c} g_{m,i}^{\mathrm{c},b_k} 
+ \sum\limits_{s \in \mathcal{C}_m} p_{s}^\mathrm{c} g_{s,i}^{\mathrm{c},b_k}
\end{align}
is the overall interference power at the receiver of the target D2D link. The variables $g_{m,j,i}^{b_k}$, $g_{s,j,i}^{b_k}$, $g_{m,i}^{\mathrm{c},b_k}$, and $g_{s,i}^{\mathrm{c},b_k}$ represent the channel power gains of $h_{m,j,i}^{b_k}$, $h_{s,j,i}^{b_k}$, $h_{m,i}^{\mathrm{c},b_k}$, and $h_{s,i}^{\mathrm{c},b_k}$, respectively. Considering the significant difference between the coverage of cellular and D2D links, the smaller maximum transmit power for a D2D link, and the total power constraint for D2D links, it is assumed that cellular communications are the major interferer of D2D communications. The assumption is appropriate in an urban scenario with dense microcells and small microcell radius. The interference is in turn modeled as a Gaussian random variable\cite{YZhao2017}.     

The corresponding energy efficiency is given by
\begin{align}\label{e:energyefficiency}
e_{m,i}(p_{m,i}) = \frac{1}{p^0_{m,i} + p_{m,i}}\log \bigg(1 + \frac{g_{m,i}^{b_k} p_{m,i}}{\sigma_0^2 + p_{m,i}^{I} }\bigg),
\end{align}    
where $p^0_{m,i}$ represents circuit power consumption. Although the energy efficiency function is nonconvex, it is unimodal and concave in $[0, p_{m,i}^*]$, where $p_{m,i}^*$ maximizes $e_{m,i}$.

With the above setup, each microBS can independently coordinate its D2D links. Accordingly, we can focus on just one microcell. The microcell index $m$ is neglected when appropriate. Correspondingly, $p_{m,i}$, $r_{m,i}$, and $e_{m,i}$ will be denoted as $p_i$, $r_i$, and $e_i$, respectively. 

As mentioned in Section~\ref{s:Sys}, each link determines its own objective $o_i$, which can be either rate or energy efficiency in this case, without needing to notify the microBS. Correspondingly, $b_i(x_i)$ in Section~\ref{s:Sys} becomes $r_i(p_i)$ if link $i$ has a rate objective or $e_i(p_i)$ otherwise. The value of a specific rate $r_i$ or energy efficiency $e_i$ to link $i$, depending on the objective, is determined by $v_i(r_i)$ or $v_i(e_i)$.

A mircoBS may coordinate its D2D links to maximize the sum-valuation, i.e,
\begin{subequations}\label{e:NUM} 
	\begin{align}
	& \mathop{\mathbf{max}}\limits_{\{p_i, i\in \mathcal{L}_m(b_k)\}} \quad \sum\limits_{i\in \mathcal{L}_m(b_k)} v_i(p_i)\\
	& \mathbf{\qquad s.t.}\quad  0 \leq p_i \leq p^{\max}, \forall i \in \mathcal{L}_m(b_k) \label{e:NUMcon1} \\
	& \qquad\quad\; \sum\limits_{i \in \mathcal{L}_m(b_k)} p_i < P^{\max} \label{e:NUMcon2}. 
	\end{align}
\end{subequations}
The above problem could accommodate more constraints, e.g., a minimum rate constraint for D2D links with an energy efficiency objective, while the results in previous sections can still apply. Nevertheless, the above basic form is used since the focus is on incentive instead of optimization. Problem \eqref{e:NUM} corresponds to the UCP, i.e., problem \eqref{e:PScenarioI}. However, when the D2D links adopt the solution to \eqref{e:NUM}, it is possible that the microBS has preference for a power allocation solution to a different problem, e.g.,
\begin{subequations}\label{e:NUMcenter} 
 	\begin{align}
 	& \mathop{\mathbf{max}}\limits_{\{p_i, i\in \mathcal{L}_m(b_k)\}} \quad \nu(\mathbf{p})\\
 	& \mathbf{\qquad s.t.}\quad  0 \leq p_i \leq p^{\max}, \forall i \in \mathcal{L}_m(b_k) \label{e:NUMcentercon1} \\
 	& \qquad\quad\; \sum\limits_{i \in \mathcal{L}_m(b_k)} p_i < P^{\max} \label{e:NUMcentercon2}. 
 	\end{align}
\end{subequations}
where $\nu(\mathbf{p})$ with $\mathbf{p}= [p_1, \dots, p_{|\mathcal{L}_m(b_k)|}]$ is the microBS's valuation representing a network preference. Problem \eqref{e:NUMcenter} corresponds to the NCP, i.e., problem \eqref{e:PScenarioII}. 

The next section will demonstrate the results obtained in the UCP and NCP for this application using the above models.

\vspace{2mm}
\section{Simulation Examples}\label{s:sim}

\begin{figure}
	\begin{center}
		\includegraphics[scale = 0.58, angle = 0]{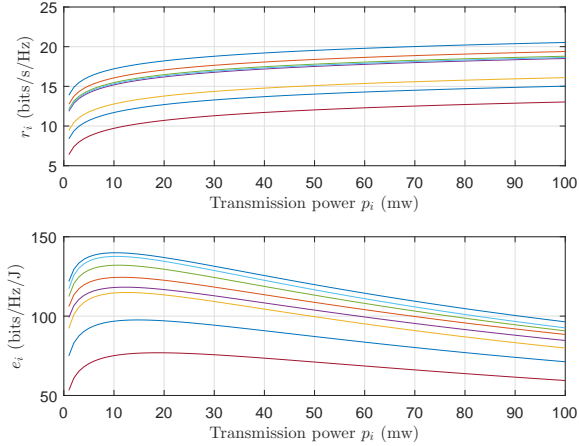}
		\caption{Transmission rates and energy efficiency versus transmit power when 8 D2D links reuse the same RB in a microcell.}
		\label{f:rateEE}
	\end{center}
	\vspace{-4mm}
\end{figure}

With the D2D application as an example, this section demonstrates the analysis on dual pricing as well as the proposed SEM and ESEM in the preceding sections. The common setup for all simulation examples are as follows. The radius of a cell is 500 meters, and the distance between D2D transceivers is in [5,25] meters. The noise spectral density is -174dB/Hz, and noise figure is 6dB. The bandwidth of an RB is 15kHz. The maximum D2D transmit power $p_i^{\max}$ is 100mw and the maximum cellular transmit power is 500mw (27dBm). The close-in (CI) model is used for Large-scale propagation \cite{SSun2016}, where non-line-of-sight (NOS) channels are assumed.\footnote{For D2D links and intra-cell interfering links, the urban micro-cellular street canyon scenario is adopted. For inter-cell interfering links, the urban macro-cellular scenario is adopted.} The small-scale fading is modeled by a Rayleigh fading channel with the channel response following $\mathcal{CN}(0,1)$. The valuation function used as an example here is $v_i(b_i) = 1 - e^{-\varepsilon_i b_i}$ \cite{CJiang2017}, where $b_i$ can be $r_i$ (given by equation \eqref{e:rate}) or $e_i$ (given by equation \eqref{e:energyefficiency}).  Fig.~\ref{f:rateEE} demonstrates the transmission rate and energy efficiency versus transmit power when eight D2D links reuse an RB in a microcell in the aforementioned setup.

%
%

\begin{figure}
	\begin{center}
		\includegraphics[scale = 0.58, angle = 0]{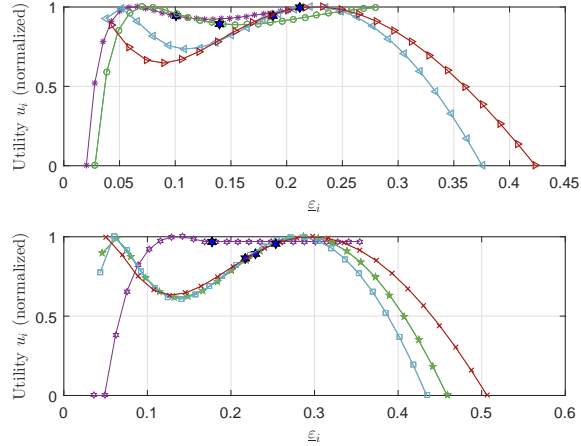}
		\caption{Untruthful reporting versus the resulting utility. Each curve corresponds to a specific user reporting untruthfully on $\varepsilon_i$ while others report truthfully.}
		\label{f:UTR_utility}
	\end{center}
	\vspace{-4mm}
\end{figure}

\begin{figure}
	\begin{center}
		\includegraphics[scale = 0.58, angle = 0]{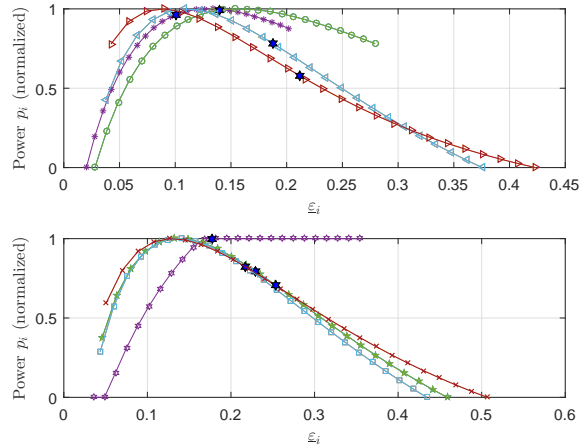}
		\caption{Untruthful reporting versus the resulting power allocation. Each curve corresponds to a specific user reporting untruthfully on $\varepsilon_i$ while others report truthfully.}
		\label{f:UTR_power}
	\end{center}
	\vspace{-4mm}
\end{figure}

\textit{ Example 1}: An illustration of the results regarding dual pricing. Given $v_i(b_i) = 1 - e^{-\varepsilon_i b_i}$, it is assumed that links report based on a truthful $\varepsilon_i$ when reporting truthfully. Eight curves, each corresponding to one D2D link, are shown in Figs.~\ref{f:UTR_utility}~and~\ref{f:UTR_power}. The curve for each link $i$ represents the resulting utility (Fig.~\ref{f:UTR_utility}) and power allocation (Fig.~\ref{f:UTR_power}) under dual pricing versus the  $\underline{\varepsilon}_i$ used by link $i$ for reporting when other links report truthfully, i.e., $\underline{\varepsilon}_j = \varepsilon_j, \forall j \neq i$.~\footnote{Each curve in Figs.~\ref{f:UTR_utility}~and~\ref{f:UTR_power} is supposed to be displayed in a separate plot. They are packed into two plots due to the limit of space. Each curve is normalized by its maximum.} The solid hexagram marker on each curve marks the utility (Fig.~\ref{f:UTR_utility}) and power allocation (Fig.~\ref{f:UTR_power}) when the corresponding link also reports truthfully, i.e., $\underline{\varepsilon}_i = \varepsilon_i$. From Fig.~\ref{f:UTR_utility}, it can be seen that reporting truthfully does not yield the maximum utility for a link. This illustrates the fact that reporting truthfully is not EPIC, as suggested by Theorem~1. Fig.~\ref{f:UTR_utility} and Fig.~\ref{f:UTR_power} jointly show that any utility larger than the utility given by truthful reporting corresponds to an allocated power smaller than the allocated power in the case of truthful reporting. This validates the result in Lemma~3.

\begin{figure}
	\begin{center}
		\includegraphics[scale = 0.58, angle = 0]{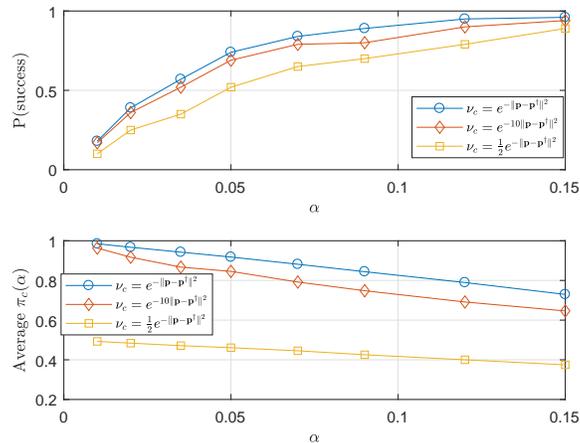}
		\caption{(a) Probability of successful exchange and (b) average utility gain from an exchange for a microBS versus  $\alpha$.}
		\label{f:SEM}
	\end{center}
	\vspace{-4mm}
\end{figure}

\textit{Example 2}: An illustration of SEM for two D2D links. In this simulation, link 1 adopts a valuation based on energy efficiency, and link 2 adopts a valuation based on rate. The parameter $\varepsilon_i$ is generated from a uniform distribution in $[0.1, 0.3]$ for each link. The valuation of the network center used as as example here is $\nu(\mathbf{p})=  a e^{- \|\mathbf{p} - \mathbf{p}^\dagger\|^2/\sigma}$. The overall interference power from all sources at each D2D receiver is generated randomly in the range of 5dB to 20dB of noise power. For each choice of the parameter $\alpha$ in SEM, 100 samples are used. The first plot in Fig.~\ref{f:SEM} shows the chance of successful exchange versus the center's choice of $\alpha$. The second plot in Fig.~\ref{f:SEM} shows the average utility gain from an exchange for a microBS versus  $\alpha$. The figure illustrates the aforementioned trade-off for the center, i.e., a larger $\alpha$ corresponds to a larger chance of success but a smaller utility gain.

%
%

\begin{figure}
	\begin{center}
	    \includegraphics[scale = 0.58, angle = 0]{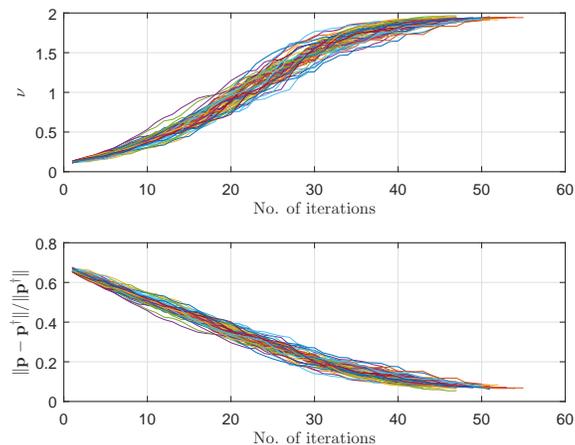}
		\caption{Traces of $\nu$ and $\|\mathbf{p} - \mathbf{p}^\dagger\|$ in 50 runs of ESEM ($a = 2$).}
		\label{f:ESEMTrace1}
	\end{center}
	\vspace{-4mm}
\end{figure}

\begin{figure}
	\begin{center}
		\includegraphics[scale = 0.47, angle = 0]{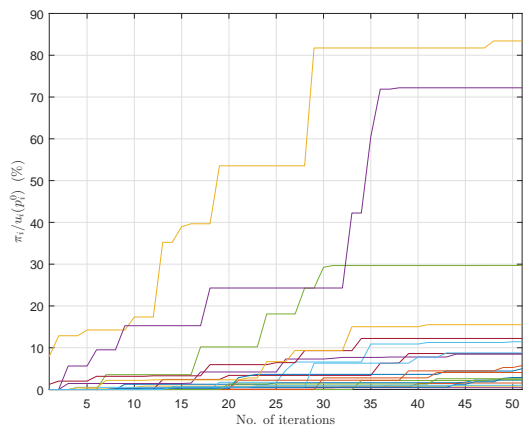}
		\caption{Traces of $\pi_i, \forall i$ in one run of ESEM ($a = 2$).}
		\label{f:ESEMTracepi}
	\end{center}
	\vspace{-4mm}
\end{figure}

\textit{Example 3}: Illustration of ESEM. In this simulation, 20 D2D links are used, five of which use energy efficiency as their objective, and the rest use data rate as their objective. The parameter $\varepsilon_i$ is randomly generated from a uniform distribution in $[0.1, 0.3]$. The valuation of the microBS used as an example is $\nu = a \exp(-\|\mathbf{p} - \mathbf{p}^\dagger\|/\sigma)$. The use of such a valuation implies that the solution found via the ESEM is compared to the optimal solution $\mathbf{p}^\dagger$, which would be achieved by a virtual optimal mechanism that is efficient, budget-balanced, and incentive compatible.  
The ESEM is implemented 50 times with $a = 2, \sigma = 0.01$. The traces of $\nu$ and $\|\mathbf{p} - \mathbf{p}^\dagger\|^2$ are shown in Fig~\ref{f:ESEMTrace1}. A step size of $\delta_l = 10^{-2}$ is used. Due to the randomness in choosing a candidate link in the ESEM, the traces from different runs can be different. In Fig~\ref{f:ESEMTrace1}, the utility gain for the microBS from an exchange is limited due to a small $a$ and, as a result, it cannot provide enough subsidy for $\mathbf{p}$ to converge to $\mathbf{p}^\dagger$ (this can be seen from the bottom subplot in Fig~\ref{f:ESEMTrace1}). However, despite of the difference in the traces, $\nu$ and $\|\mathbf{p} - \mathbf{p}^\dagger\|$ at the output are almost identical for each run. For example, the average $\nu$ at the final iteration in Fig~\ref{f:ESEMTrace1} over the 50 runs is 1.9695 while the difference between the maximum and the minimum $\nu$ is only $3.1\%$ of this average. \footnote{With a larger choice of $a$, $\mathbf{p}$ can converge to $\mathbf{p}^\dagger$. But this result is omitted due to limit of space.}. 
The progress of utility gain for all 20 D2D links in one run of ESEM is shown in Fig~\ref{f:ESEMTracepi}. It can be seen that the utilities of all links are nondecreasing in ESEM, which illustrates the individual rationality property of ESEM. 

\section{Conclusion}

Incentive for truthful reporting is studied for utility optimization problems with local information. The proposed model allows each user to adopt an individual objective and a valuation function, incorporated as its local information, and connects the optimization and the incentive perspectives. 
Dual pricing is shown to be not EPIC for finding a solution to the UCP. With dual pricing, reporting truthfully is shown to be optimal for a user only when the resource is over-supplied or when the user cannot afford any resource. If the network center intends to adopt a solution different from the one accepted by the users, the solution needs to be also preferred by at least one user. Otherwise, no incentive mechanism can specify a condition to guarantee the implementation of the solution and at the same time guarantee truthful reporting. In the two-user case, whether such a solution can be implemented is determined in one shot. In the multiuser case, the switch between solutions must go through an iterative process. The proposed SEM and ESEM achieve incentive compatibility and provide nonnegative utility gains to both the users and the center. Simulations for a D2D application scenario validate the analysis of dual pricing and demonstrate the effectiveness of the proposed SEM and ESEM.

\appendix

\subsection{Proof of Lemma~1}\label{p:Lemma1}

When $\mathrm{s}_i = \mathrm{s}_i^\mathrm{TR}$, $\underline{x}_i^\star = x_i^\star$ and $\underline{\lambda}^\star = \lambda^\star$. From the primal optimality, the following result holds:
\begin{align}
&u_i(\{\mathrm{s}_i^\mathrm{TR}, \mathrm{s}_{-i}^\mathrm{TR}\}) = v_i(x_i^\star) -\lambda^\star x_i^\star  = \sup\limits_{x_i}  \big(v_i(x_i) - \lambda^\star x_i\big) \nonumber \\ 
&\;\; \geq  v_i(\underline{x}_i^\star) - \lambda^\star \underline{x}_i^\star 
=  v_i(\underline{x}_i^\star) - \underline{\lambda}^\star \underline{x}_i^\star - (\lambda^\star - \underline{\lambda}^\star)\underline{x}_i^\star
\end{align}
where $\underline{\lambda}^\star$ and $\underline{x}_i^\star$ are the final price and the allocation to user $i$ when $\mathrm{s}_i \neq \mathrm{s}_i^\mathrm{TR}$, respectively.
If $\lambda^\star < \underline{\lambda}^\star$, i.e., user $i$ unilaterally drives the price higher, then it follows that $- (\lambda^\star - \underline{\lambda}^\star)\underline{x}_i^\star \geq 0$ and
\begin{align}
u_i(\{\mathrm{s}_i^\mathrm{TR}, \mathrm{s}_{-i}^\mathrm{TR}\}) & =  v_i(x_i^\star) -\lambda^\star x_i^\star  \nonumber \\ & \geq v_i(\underline{x}_i^\star) - \underline{\lambda}^\star \underline{x}_i^\star = \underline{u}_i(\{\mathrm{s}_i, \mathrm{s}_{-i}^\mathrm{TR}\}),
\end{align}
where the strict equality holds when $\underline{x}_i^\star > 0$. Accordingly, user $i$ achieves a less utility with such $\mathrm{s}_i \neq \mathrm{s}_i^\mathrm{TR}$ that renders $\underline{\lambda}^\star(\{\mathrm{s}_i, \mathrm{s}_{-i}^\mathrm{TR}\}) > \lambda^\star(\{\mathrm{s}_i^\mathrm{TR}, \mathrm{s}_{-i}^\mathrm{TR}\})$. \hfill $\blacksquare$

\subsection{Proof of Lemma~2}\label{p:Lemma2}

Define  $\mathcal{D}_i = [0, \infty)$ if $v_i(x_i)$ is nondecreasing and concave, and $\mathcal{D}_i = [0, x_i^*]$ if $v_i(x_i)$ is unimodal and concave in $[0, x_i^*]$, where $x_i^*$ maximizes $v_i(x_i)$. For any given $v_i(x_i)$ that satisfies the aforementioned properties (i.e., nondecreasing and concave in $\mathcal{D}_i$, twice-differentiable, and  $v_i(0) = 0$), a $\underline{v}_i(x_i)$ can be found such that $\underline{v}_i(x_i) < v_i(x_i), \forall x_i \in \mathcal{D}_i$ and $\underline{v}_i^\prime(x_i) < v_i^\prime(x_i), \forall x_i \in \mathcal{D}_i$. \footnote{A simple example of such reporting strategy is using $\underline{v}_i = \alpha v_i$ where $\alpha \in(0,1)$. } Denote strategy $\mathrm{s}_i \neq \mathrm{s}_i^\mathrm{TR}$ as the strategy of reporting based on $\underline{v}_i$. Four cases are possible.

\textit{Case i}: $\underline{v}^\star_j(\mathrm{s}_i, \mathrm{s}_{-i}^\mathrm{TR}) = v^\star_j(\mathrm{s}_i^\mathrm{TR}, \mathrm{s}_{-i}^\mathrm{TR}) =  0, \forall j\neq i$. In this case, the following result holds:
\begin{align}
\underline{\lambda}^\star(\mathrm{s}_i, \mathrm{s}_{-i}^\mathrm{TR}) \!\stackrel{(a)}{=}\!  \underline{v}_i^\prime(\underline{x}^\star_i) \! \stackrel{(b)}{<} \! v_i^\prime(\underline{x}^\star_i) \!=\! v_i^\prime(x^\star_i) \!\stackrel{(a)}{=}\! \lambda^\star(\mathrm{s}_i^\mathrm{TR}, \mathrm{s}_{-i}^\mathrm{TR}),
\end{align}
where the equalities with mark $(a)$ follow from KKT conditions, and the inequality with mark $(b)$ holds because $\underline{v}_i^\prime(x_i) < v_i^\prime(x_i), \forall x_i$.~\footnote{The same meanings apply for notations on the equality and inequality signs, respectively, in the rest of the proof.} In this case, $\underline{x}^\star_i(\mathrm{s}_i,  \mathrm{s}_{-i}^\mathrm{TR}) = x^\star_i(\mathrm{s}_i^\mathrm{TR}, \mathrm{s}_{-i}^\mathrm{TR}) = X^{\max}$.

\textit{Case ii}: $\exists j\neq i$ such that $\underline{x}^\star_j(\mathrm{s}_i, \mathrm{s}_{-i}^\mathrm{TR}) > x^\star_j(\mathrm{s}_i^\mathrm{TR}, \mathrm{s}_{-i}^\mathrm{TR}) = 0$. In this case, the following result holds:
\begin{align}
\underline{\lambda}^\star(\mathrm{s}_i, \mathrm{s}_{-i}^\mathrm{TR}) \!\stackrel{(a)}{=}\!  v_j^\prime(\underline{x}^\star_j) \stackrel{(c)}{<} v_j^\prime(0) \!\leq\! v_i^\prime(x^\star_i) \!\stackrel{(a)}{=}\! \lambda^\star(\mathrm{s}_i^\mathrm{TR}, \mathrm{s}_{-i}^\mathrm{TR}),
\end{align}
where the inequality with mark $(c)$ holds due to $\underline{v}_j(x_j)$ being strictly concave, and $v_j^\prime(0) \leq v_i^\prime(x^\star_i)$ holds because $x^\star_j$ must be larger than zero and must satisfy $ v_i^\prime(x^\star_i) =  v_j^\prime(x^\star_j)$ otherwise. In this case, $\underline{x}^\star_i(\mathrm{s}_i,  \mathrm{s}_{-i}^\mathrm{TR}) < x^\star_i(\mathrm{s}_i^\mathrm{TR}, \mathrm{s}_{-i}^\mathrm{TR}) = X^{\max}$.

\textit{Case iii}: $\exists j\neq i$ such that $x^\star_j(\mathrm{s}_i^\mathrm{TR}, \mathrm{s}_{-i}^\mathrm{TR}) > \underline{x}^\star_j(\mathrm{s}_i, \mathrm{s}_{-i}^\mathrm{TR}) = 0$. This case is impossible because it leads to the following result:
\begin{align}
\underline{\lambda}^\star(\mathrm{s}_i, \mathrm{s}_{-i}^\mathrm{TR}) \! \stackrel{(a)}{=} \!   \underline{v}_i^\prime(\underline{x}^\star_i)  \!\geq\! v_j^\prime(0)  &\!\stackrel{(c)}{>}\! v_j^\prime(x^\star_j) \nonumber \\ & \!\stackrel{(a)}{=}\! v_i^\prime(x^\star_i) \!\stackrel{(a)}{=}\! \lambda^\star(\mathrm{s}_i^\mathrm{TR}, \mathrm{s}_{-i}^\mathrm{TR}).
\end{align}
The result $\underline{\lambda}^\star(\mathrm{s}_i, \mathrm{s}_{-i}^\mathrm{TR}) > \lambda^\star(\mathrm{s}_i^\mathrm{TR}, \mathrm{s}_{-i}^\mathrm{TR})$ infers $\sum_{j\neq i} \underline{x}^\star_j < \sum_{j\neq i} x^\star_j$. The result $\underline{v}_i^\prime(\underline{x}^\star_i) > v_i^\prime(x^\star_i)$ infers $\underline{x}^\star_i < x^\star_i$ since $\underline{v}_i^\prime(x_i) < v_i^\prime(x_i), \forall x_i \in \mathcal{D}_i$. As a result, $\underline{x}^\star_i + \sum_{j\neq i} \underline{x}^\star_i < x^\star_i + \sum_{j\neq i} x^\star_j \leq X^{\max}$, which occurs only if $\underline{\lambda}^\star(\mathrm{s}_i, \mathrm{s}_{-i}^\mathrm{TR}) = 0$ based on the KKT conditions. It contradicts with  $\underline{\lambda}^\star(\mathrm{s}_i, \mathrm{s}_{-i}^\mathrm{TR}) > \lambda^\star(\mathrm{s}_i^\mathrm{TR}, \mathrm{s}_{-i}^\mathrm{TR})$. Therefore, this case is impossible.

\textit{Case iv}: $\underline{x}^\star_j(\mathrm{s}_i, \mathrm{s}_{-i}^\mathrm{TR}) > 0$ and $x^\star_j(\mathrm{s}_i^\mathrm{TR}, \mathrm{s}_{-i}^\mathrm{TR}) > 0, \forall j$. In this case, the effect of user $i$ adopting $\mathrm{s}_i$ instead of $\mathrm{s}_i^\mathrm{TR}$ can be shown by contradiction. Suppose that $\underline{\lambda}^\star (\mathrm{s}_i, \mathrm{s}_{-i}^\mathrm{TR}) \geq \lambda^\star(\mathrm{s}_i^\mathrm{TR}, \mathrm{s}_{-i}^\mathrm{TR})$. Then, it follows that:
\begin{align}
\!\!v_j^\prime(\underline{x}_j^\star) \!\!\stackrel{(a)}{=}\!\! \underline{\lambda}^\star(\mathrm{s}_i, \mathrm{s}_{-i}^\mathrm{TR})  & \!\geq\!\! \lambda^\star(\mathrm{s}_i^\mathrm{TR}\!, \mathrm{s}_{-i}^\mathrm{TR}) \!\!\stackrel{(a)}{=}\!\!  v_j^\prime(x^\star_j) \!\!\stackrel{(a)}{=}\!\!  v_i^\prime(x^\star_i) \!\! > \!0, \label{e:Lemma2Cond1}\\
\!\!\underline{\lambda}^\star(\mathrm{s}_i, \mathrm{s}_{-i}^\mathrm{TR})  & \stackrel{(d)}{\geq}  \underline{v}_i^\prime(\underline{x}^\star_i),  \label{e:Lemma2Cond2}
\end{align}
where the strict equality holds in step $(d)$ if $\underline{x}^\star_i(\mathrm{s}_i, \mathrm{s}_{-i}^\mathrm{TR}) \neq 0$. From \eqref{e:Lemma2Cond1}, it follows that $\underline{x}^\star_j(\mathrm{s}_i, \mathrm{s}_{-i}^\mathrm{TR}) \leq x^\star_j(\mathrm{s}_i^\mathrm{TR}, \mathrm{s}_{-i}^\mathrm{TR}), \forall j$. If the equality in step $(d)$ applies, then $\underline{v}_i^\prime(\underline{x}^\star_i) \geq v_i^\prime(x^\star_i)$ based on \eqref{e:Lemma2Cond1} and \eqref{e:Lemma2Cond2}. Using the property  $\underline{v}_i^\prime(x_i) < v_i^\prime(x_i), \forall x_i$, it follows that $\underline{v}_i^\prime(x^\star_i) < v_i^\prime(x^\star_i) \leq \underline{v}_i^\prime(\underline{x}^\star_i)$.  Since $\underline{v}_i$ is strictly concave, it follows that $\underline{x}^\star_i(\mathrm{s}_i, \mathrm{s}_{-i}^\mathrm{TR}) < x^\star_i(\mathrm{s}_i^\mathrm{TR}, \mathrm{s}_{-i}^\mathrm{TR})$. If the strict inequality in step $(d)$ applies, if follows that:
\begin{align}
\underline{x}^\star_i(\mathrm{s}_i, \mathrm{s}_{-i}^\mathrm{TR}) = 0 < x^\star_i(\mathrm{s}_i^\mathrm{TR}, \mathrm{s}_{-i}^\mathrm{TR})
\end{align}
as $x^\star_i(\mathrm{s}_i^\mathrm{TR}, \mathrm{s}_{-i}^\mathrm{TR}) \neq 0$ in the lemma. Therefore, $\underline{x}^\star_i(\mathrm{s}_i, \mathrm{s}_{-i}^\mathrm{TR}) < x^\star_i(\mathrm{s}_i^\mathrm{TR}, \mathrm{s}_{-i}^\mathrm{TR})$ in either case. Since $\underline{x}^\star_j \leq x^\star_j, \forall j$, it leads to the result that
\begin{align}
\underline{x}^\star_i + \sum_{j\neq i} \underline{x}^\star_i < x^\star_i + \sum_{j\neq i} x^\star_j \leq X^{\max},
\end{align}
which happens only if $\underline{\lambda}^\star(\mathrm{s}_i, \mathrm{s}_{-i}^\mathrm{TR}) = 0$. It contradicts with  $\underline{\lambda}^\star(\mathrm{s}_i, \mathrm{s}_{-i}^\mathrm{TR}) \geq \lambda^\star(\mathrm{s}_i^\mathrm{TR}, \mathrm{s}_{-i}^\mathrm{TR}) >0$.  Therefore, the assumption $\underline{\lambda}^\star (\mathrm{s}_i, \mathrm{s}_{-i}^\mathrm{TR}) \geq \lambda^\star(\mathrm{s}_i^\mathrm{TR}, \mathrm{s}_{-i}^\mathrm{TR})$ must be invalid, which proves $\underline{\lambda}^\star (\mathrm{s}_i, \mathrm{s}_{-i}^\mathrm{TR}) < \lambda^\star(\mathrm{s}_i^\mathrm{TR}, \mathrm{s}_{-i}^\mathrm{TR})$ in this case.

Summarizing the above four cases concludes the proof. \hfill $\blacksquare$

\subsection{Proof of Lemma~3}\label{p:Lemma3}



Since it is impossible that $x_l^\star = 0, \forall l$, there must exist at least one user, denoted as user $i$, and a corresponding $\mathrm{s}_i \neq \mathrm{s}_i^\mathrm{TR}$ so that $\underline{\lambda}^\star(\mathrm{s}_i, \mathrm{s}_{-i}^\mathrm{TR})  < \lambda^\star(\mathrm{s}_i^\mathrm{TR}, \mathrm{s}_{-i}^\mathrm{TR})$ according to Lemma 2. Moreover, based on the four cases in the proof of Lemma 2, the impact of $\mathrm{s}_i$ satisfies $\underline{x}^\star_i \leq x^\star_i$ in all possible cases. Given that $\underline{\lambda}^\star < \lambda^\star$, $\underline{x}_j^\star \geq x_j^\star, \forall j\neq i$ and the strict inequality holds if $\underline{x}_j^\star \neq 0$. Two cases are possible. 

\textit{Case i}: $\max\limits_j v_j^\prime(0) < \lambda^\star$.  As a result, $\exists \mathrm{s}_i$ such that $\underline{\lambda}^\star \in [\max\limits_j v_j^\prime(0), \lambda^\star)$,  $\underline{x}_j^\star = 0, \forall j \neq i$, and $\underline{x}_i^\star = x_i^\star$. Accordingly, the following result holds
\begin{align}
\underline{u}_i(\mathrm{s}_i, \mathrm{s}_{-i}^\mathrm{TR}) = v_i(\underline{x}_i^\star) - \underline{\lambda}^\star \underline{x}_i^\star = v_i(x_i^\star) - \underline{\lambda}^\star x_i^\star  \nonumber \\ > v_i(x_i^\star) - \lambda^\star x_i^\star = u_i(\mathrm{s}_i^\mathrm{TR}, \mathrm{s}_{-i}^\mathrm{TR}).
\end{align}

\textit{Case ii}: $\max\limits_j v_j^\prime(0) \geq \lambda^\star$. As a result, $\exists j \neq i$ such that $\underline{x}_j^\star > x_j^\star, \forall \underline{\lambda}^\star(\mathrm{s}_i, \mathrm{s}_{-i}^\mathrm{TR}) < \lambda^\star$. Define
\begin{subequations}
\begin{align}
\mathcal{I}_{-i} = \{j| j \in \mathcal{U}/\{i\}, \underline{x}_j^\star > x_j^\star\}, \\
\mathcal{U}_{-i} = \{j| j \in \mathcal{U}/\{i\}, \underline{x}_j^\star = x_j^\star\}.
\end{align}
\end{subequations}
Since  $\underline{x}_j^\star \geq x_j^\star, \forall j\neq i$, $\mathcal{I}_{-i}\cup  \mathcal{U}_{-i} = \mathcal{U}/\{i\}$. Given $\underline{x}_i^\star < x_i^\star$, $\underline{x}_j^\star > x_j^\star$ for $j \in \mathcal{I}_{-i}$, and $\underline{x}_j^\star = x_j^\star$ for $j \in \mathcal{U}_{-i}$, denote  $\underline{x}_i^\star = x_i^\star - \Delta x_i $ and, accordingly, $\underline{x}_j^\star = x_j^\star + \Delta x_j, \forall j\in \mathcal{I}_{-i}$, where $\Delta x_i  > 0, \Delta x_j > 0, \Delta x_i = \sum\limits_{j\in \mathcal{I}_{-i}}\Delta x_j$. Pick any $j$ with $\lambda^\star = v_j^\prime(x_j^\star)$ from $\mathcal{I}_{-i}$. Suppose that Lemma~3 is not true. The following result must hold:
\begin{align}
u_i(\mathrm{s}_i^\mathrm{TR}, \mathrm{s}_{-i}^\mathrm{TR}) 
& = v_i(x_i^\star) \!-\! v_j^\prime(x^\star_j) x_i^\star  \nonumber \\ 
& \!\geq\!  v_i(x_i^\star \!-\! \Delta x_i) \!-\! v_j^\prime(x_j^\star \!+\! \Delta x_j) (x_i^\star \!-\! \Delta x_i) \nonumber \\
& = \underline{u}_i(\mathrm{s}_i, \mathrm{s}_{-i}^\mathrm{TR}) \label{e:lemma3Ui}
\end{align}
for all possible $\Delta x_i$ such that $\underline{x}_i^\star = x_i^\star - \Delta x_i$ falls in $\mathcal{D}_i$. Let $\Delta x_i \rightarrow 0$, reformat \eqref{e:lemma3Ui}, and take limit on both sides. The following result holds:
\begin{align}
&\lim\limits_{\Delta x_i \rightarrow 0} \!\! \frac{ v_i(x_i^\star) \!-\! v_i(x_i^\star \!-\! \Delta x_i)}{\Delta x_i} \nonumber \\ & \geq \!\!  \lim\limits_{\Delta x_i \rightarrow 0} \bigg(\!\!x_i^\star \frac{ v_j^\prime(x^\star_j)  \!-\! v_j^\prime(x_j^\star \!+\! \Delta x_j)}{\Delta x_j} \frac{\Delta x_j}{\Delta x_i}  \!+ \! v_j^\prime(x_j^\star \!+\! \Delta x_j) \bigg). \label{e:lemma3UiRes1}
\end{align}
Since $v_i$ are twice differentiable in $\mathcal{D}_i$, the above result can be rewritten as 
\begin{align}
v_i^\prime(x_i^\star) \geq - v_j^{\prime\prime}(x^\star_j)x_i^\star \gamma_{ji} + v_j^\prime(x_j^\star) > v_j^\prime(x_j^\star), \label{e:lemma3UiRes2}
\end{align}
where $\gamma_{ji} = \lim\limits_{\Delta x_i \rightarrow 0}{\Delta x_j}/{\Delta x_i}$. The second inequality is due to the fact that $v_i, \forall i$ is strictly concave, and $x_i^\star$ and $\gamma_{ji}$ are both positive values. However, \eqref{e:lemma3UiRes2} contradicts with $v_i^\prime(x_i^\star) = \lambda^{\star} = v_j^\prime(x_j^\star)$. 
%

This completes the proof of Lemma~3. \hfill$\blacksquare$


\subsection{Proof of Lemma~4}\label{p:Lemma4}

A mechanism is either one-shot or iterative. The following cases are possible.

\textit{Case i}:  one-shot, center initiates. In this case, center proposes a transfer $t_i$, user $i$ decides to accept it or not based on the comparison between $v_i(x_i^\star) - v_i(x_i^\dagger)$ and $t_i$. Truthful reporting is optimal for user $i$. However, since the center has no information on $v_i$, it is impossible to guarantee $ t_i > v_i(x_i^\star) - v_i(x_i^\dagger)$ in one shot. Consequently, no condition can be found to guarantee that $\nu(\mathbf{x}^\dagger) - \nu(\mathbf{x}^\star) \geq \sum_i t_i $. 

\textit{Case ii}: one-shot, user initiates. In this case, user $i$ proposes a $t_i$ based on $v_i(x_i^\dagger) - v_i(x_i^\star)$, center decides to accept it or not based on $s_c - \sum\limits_i t_i = \nu(\mathbf{x}^\dagger) - \nu(\mathbf{x}^\star) - \sum\limits_i t_i$. Truthful reporting is optimal for user $i$ only if the maximum $t_i$ acceptable to the center, denoted as $t^\mathrm{a}_i$, satisfies $t^\mathrm{a}_i \leq v_i(x_i^\star) - v_i(x_i^\dagger)$. However, $\mathbf{x}^\dagger$ cannot be implemented if $t^\mathrm{a}_i < v_i(x_i^\star) - v_i(x_i^\dagger)$  while $t_i$ happens to equal $t^\mathrm{a}_i$ in one shot is impossible given that user $i$ has no information on $\nu$.

\textit{Case iii}: iterative, center initiates. In such case, truthful reporting is optimal only if the center uses a fixed total $t^{T} = \sum\limits_i t_i = \sum\limits_i (v_i(x_i^\star) - v_i(x_i^\dagger))$ and adjusts the division of $t^{T}$ to $t_1$ and $t_2$ through the iterations. If the users know $t^{T} = \sum\limits_i t_i = \sum\limits_i (v_i(x_i^\star) - v_i(x_i^\dagger))$ in all iterations, truthful reporting is optimal for everyone. However, it is impossible for the center to figure out such $t^{T}$ and use it in every iteration since $v_i$ is local information. Any $t^{T} > \sum\limits_i t_i$ leads to untruthful reporting. Any $t^{T} < \sum\limits_i t_i$ cannot implement $\mathbf{x}^\dagger$. A varying $t^{T}$ across iterations leads to either untruthful reporting or fail of implementing $\mathbf{x}^\dagger$.

\textit{Case iv}: iterative, user initiates. Reporting is not truthful in such case because user $i$ proposes different $t_i$ through the iterations, which cannot all be equal to $v_i(x_i^\dagger) - v_i(x_i^\star)$.

In summary, a mechanism cannot simultaneously be incentive compatible and guarantee the implementation of $\mathbf{x}^\dagger$ when $x_i^\dagger \leq x_i^\star, \forall i$. \hfill$\blacksquare$

\subsection{Proof of Lemma~5}\label{p:Lemma5}

The utility gain for the users, as a function of quotes $\rho$ and $\varphi$, from participating in the procedure in Algorithm~\ref{a:SEM} is 
\begin{subequations}
	\begin{align} 
	\pi_1(\varphi, \rho) &= \big((\varphi + \alpha s_c) - \rho^\mathrm{TR}\big)^+ \label{e:Lemma4u1} \\
	\pi_2(\varphi, \rho) &= \big(\varphi^\mathrm{TR} - (\rho - \alpha s_c)\big)^+ \label{e:Lemma4u2}
	\end{align}
\end{subequations}
where $\rho^\mathrm{TR} = v_1(x_1^\star) - v_1(x_1^\dagger)$ and $\varphi^\mathrm{TR} = v_2(x_2^\dagger) - v_2(x_2^\star)$. The superscript $(\cdot)^+$ represents the projection onto the nonnegative orthant. The values $\rho^\mathrm{TR}$ and $\varphi^\mathrm{TR}$ represent the true cost for user 1 and the true gain of user 2 in the exchange, respectively. The projection reflects the fact that an exchange is made only if $\varphi + \alpha s_c  \geq \rho$. The reported and the true values satisfy $\rho \geq \rho^\mathrm{TR}$ and $\varphi\leq \varphi^\mathrm{TR}$ for any rational users.

DSIC is achieved if $\pi_1(\varphi, \rho)\leq \pi_1(\varphi, \rho^\mathrm{TR}), \forall \rho, \forall \varphi$ and $\pi_2(\varphi, \rho)\leq \pi_2(\varphi^\mathrm{TR}, \rho), \forall \varphi, \forall \rho$. From \eqref{e:Lemma4u1} and \eqref{e:Lemma4u2}, it can be seen that $\pi_1(\varphi, \rho)$ is in fact independent on $\rho$, and $\pi_2(\varphi, \rho)$ is independent on $\varphi$. Therefore, when $\varphi + \alpha s_c  \geq \rho$, it holds that $\pi_1(\varphi, \rho)= \pi_1(\varphi, \rho^\mathrm{TR}), \forall \varphi$ and $\pi_2(\varphi, \rho) = \pi_2(\varphi^\mathrm{TR}, \rho), \forall \rho$. Meanwhile, from the perspective of user 1, the chance that $\varphi + \alpha s_c  \geq \rho$ is smaller than the chance that $\varphi + \alpha s_c  \geq \rho^\mathrm{TR}$ for any $\rho>\rho^\mathrm{TR}$. Therefore, reporting a $\rho$ with $\rho > \rho^\mathrm{TR}$ does not increase $\pi_1(\varphi, \rho)$ when an exchange is made but increases the chance of no exchange, in which case $\pi_1(\varphi, \rho) = 0 \leq \pi_1(\varphi, \rho^\mathrm{TR})$. Similarly, reporting a $\varphi$ with $\varphi < \varphi^\mathrm{TR}$ does not increase $\pi_2(\varphi, \rho)$ when an exchange is made but increases the chance of no exchange, in which case $\pi_2(\varphi, \rho) = 0 \leq \pi_2(\varphi^\mathrm{TR}, \rho)$. Therefore,  $\pi_1(\varphi, \rho)\leq \pi_1(\varphi, \rho^\mathrm{TR}), \forall \rho, \forall \varphi$ and $\pi_2(\varphi, \rho)\leq \pi_2(\varphi^\mathrm{TR}, \rho), \forall \varphi, \forall \rho$ regardless of whether the condition $\varphi + \alpha s_c  \geq \rho$ is satisfied or not.

Individual rationality follows from the fact that $\pi_1(\varphi, \rho) \geq 0$ and $\pi_2(\varphi, \rho) \geq 0$ based on \eqref{e:Lemma4u1} and \eqref{e:Lemma4u2}. \hfill$\blacksquare$

\subsection{Proof of Theorem~2}\label{p:Theorem2}

The proof for individual rationality follows the same logic as that in the proof of Lemma~5 and is omitted. Proof of incentive compatibility is given as follows. Truthful Report means that
\begin{subequations}
	\begin{align}
	\rho_i^l &= v_i(x_i^l) - v_i(x_i^l - \delta^l), i \in \mathrm{S}_1^l \\
	\varphi_j^l &= v_j(x_j^l + \delta^l) - v_j(x_j^l), j \in \mathrm{S}_2^l. 
	\end{align} 
\end{subequations}
We prove that it is optimal for user $i, i\in\mathrm{S}_1^l,$ to report truthfully assuming other users do the same. The proof for $j \in \mathrm{S}_2^l$ follows the same logic. Denote a report not necessarily truthful as $\underline{\rho}_i^l$.\footnote{It is assumed by default that $i \in \mathrm{S}_1^l$ in the rest of the proof.} All other users report truthfully. Denote the resulting matrix $\mathrm{W}^l$ when user $i$ reports $\underline{\rho}_i^l$ as $\mathrm{W}^l(\underline{\rho}_i^l)$. Denoted the events that user $i = \tilde{i}$ (i.e., row $i$ of $\mathrm{W}^l$ is selected in the fifth step of Algorithm~\ref{a:ESEM}) and $i \neq \tilde{i}$ with report $\underline{\rho}_i^l$ as $\mathrm{C}_i(\underline{\rho}_i^l)$ and $\mathrm{\bar{C}}_i(\underline{\rho}_i^l)$, respectively. Two cases are possible.

\textit{Case i}: $\underline{\rho}_i^l \neq \rho_i^l$ but $\mathrm{W}^l(\underline{\rho}_i^l) = \mathrm{W}^l(\rho_i^l)$. In such case,  the probabilities of $\mathrm{C}_i(\underline{\rho}_i^l)$ and $\mathrm{C}_i(\rho_i^l)$ are equal. The selection of $\tilde{j}$ given the event $\mathrm{C}_i(\underline{\rho}_i^l)$ is independent on the $\underline{\rho}_i^l$ given that $\mathrm{W}^l(\underline{\rho}_i^l)= \mathrm{W}^l(\rho_i^l)$. Moreover, the utility gain 
\begin{align}\label{e:iUGain}
\pi_i^l(\underline{\rho}_i^l) = \alpha^l \theta_{i\tilde{j}}^l + \varphi_{\tilde{j}}^l - \rho_i^l
\end{align}
is independent on $\underline{\rho}_i^l$. Therefore, $\pi_i^l(\underline{\rho}_i^l) = \pi_i^l(\rho)$ assuming $\mathrm{C}_i(\underline{\rho}_i^l)$ and $\pi_i^l(\underline{\rho}_i^l) = \pi_i^l(\rho) = 0$ assuming $\mathrm{\bar{C}}_i(\underline{\rho}_i^l)$. It can also be observed that $\underline{\rho}_i^l$ in such a case will not have an effect that propagates into any future iteration. 

\textit{Case ii}: $\underline{\rho}_i^l \neq \rho_i^l$ and $\mathrm{W}^l(\underline{\rho}_i^l) \neq \mathrm{W}^l(\rho_i^l)$. In this case, three sub-cases are possible. 

Sub-case ii-1: $i = \tilde{i}$ with both $\underline{\rho}_i^l$ and $\rho_i^l$, and $\mathrm{W}_{i, \tilde{j}}^l(\underline{\rho}_i^l) = \mathrm{W}_{i, \tilde{j}}^l(\rho_i^l)$, i.e., the selected element is not affected by untruthful report. In this sub-case, $\pi_i^l(\underline{\rho}_i^l) = \pi_i^l(\rho_i^l)$. 

Sub-case ii-2: $\mathrm{W}_{i, \tilde{j}}^l(\rho_i^l) = 1$ and $\mathrm{W}_{i, \tilde{j}}^l(\underline{\rho}_i^l) = 0$. In this sub-case,  $\pi_i^l(\underline{\rho}_i^l) = 0 \leq \pi_i^l(\rho_i^l)$. Moreover, as user $i$ makes the center believe that $\underline{\rho}_i^l$ is a truthful report, the following result holds 
\begin{align}
\psi_{i\tilde{j}}^t(\rho_i^t) & =  \varphi_{\tilde{j}}^t + \alpha^t \theta_{i\tilde{j}}^t -\rho_i^t \stackrel{(a)}{\leq} \varphi_{\tilde{j}}^l + \alpha^l \theta_{i\tilde{j}}^l - \rho_i^l \nonumber \\
&\stackrel{(b)}{=} \varphi_{\tilde{j}}^l + \alpha^l \theta_{i\tilde{j}}^l - \underline{\rho}_i^l \leq 0
\end{align}
from the perspective of the center, where step (a) is based on the facts that $\varphi_{j}^t \leq \varphi_{j}^l$, $\rho_{i}^t \geq \rho_{i}^l$, and $\alpha^t\theta_{i,\tilde{j}}^t < \alpha^l\theta_{i, \tilde{j}}^l$ when $\tilde{i} = i$, $\forall i\in \mathrm{S}_1^l \cap \mathrm{S}_1^t, \forall j \in \mathrm{S}_2^l \cap \mathrm{S}_2^t, \forall t > l$ such that $\delta^l = \delta^t$, and step (b) is based on the belief $\underline{\rho}_i^l = \rho_i^l$. \footnote{The center does not know $\rho_i^l$ but takes $\underline{\rho}_i^l$ as a truthful report and consider it to be $\rho_i^l$.} Accordingly,  $\psi_{i\tilde{j}}^t(\rho_i^t) \leq 0$ results in $\mathrm{W}_{i, \tilde{j}}^t(\rho_i^t) = 0, \forall t > l$. Consequently, user $i$ looses the opportunity to make profitable exchange with user $\tilde{j}$ in all future iterations. Therefore, any such untruthful report $\underline{\rho}_i^l$ is non-beneficial for user $i$ in the current and future iterations in this case.\footnote{This is true whenever the consistency requirement $\rho_i^t > \rho_i^l$ is satisfied. Otherwise, user $i$ exposes that it is reporting untruthfully.}

Sub-case ii-3: $\mathrm{W}_{i, \tilde{j}}^l(\rho_i^l) = 0$ and $\mathrm{W}_{i, \tilde{j}}^l(\underline{\rho}_i^l) = 1$. In this sub-case, 
\begin{align}
\pi_i^l(\underline{\rho}_i^l) = \alpha^l \theta_{i\tilde{j}}^l + \varphi_{\tilde{j}}^l - \rho_i^l < 0.
\end{align}
The impact of $\underline{\rho}_i^l$ will not propagate into any future iteration since $\mathrm{W}_{i, \tilde{j}}^{l+1}(\underline{\rho}_i^{l+1})$ will be recalculated. 

Summarizing Sub-cases ii-1 to ii-3, it can be seen that any change in an element of $\mathrm{W}(\rho_i^l)$ due to untruthful reporting leads to the same or less utility gain for user $i$ in Case ii. 

Combining Case i and Case ii proves Theorem~II. \hfill$\blacksquare$


\begin{thebibliography}{99}


\bibitem{JYJ2013}
J. Y. Joo and M. D. Ili\'{c}, ``Multi-Layered Optimization Of Demand Resources Using Lagrange Dual Decomposition,'' {\it IEEE Trans. Smart Grid}, vol.~4, no.~4, pp.~2081--2088, Dec.~2013.

\bibitem{KW2017}
K. Wada and K. Sakurama, ``Privacy Masking for Distributed Optimization and Its Application to Demand Response in Power Grids,''  {\it IEEE Trans. Ind. Electron.}, vol.~64, no.~6, pp.~5118--5128, June~2017.

\bibitem{ML2013}
M. Leinonen, M. Codreanu, and M. Juntti, ``Distributed Joint Resource and Routing Optimization in Wireless Sensor Networks via Alternating Direction Method of Multipliers,''  {\it IEEE Trans. Wireless Commun.}, vol.~12, no.~11, pp.~5454--5467, Nov.~2013.

\bibitem{TChang2014}
T. H. Chang, A. Nedi\'{c} and A. Scaglione, ``Distributed Constrained Optimization by Consensus-Based Primal-Dual Perturbation Method," {\it IEEE Trans. Autom. Control}, vol.~59, no.~6, pp.~1524--1538, June~2014.

\bibitem{DT2016}
D. Tian, J. Zhou, Z. Sheng, and V. C. M. Leung, ``Robust Energy-Efficient MIMO Transmission for Cognitive Vehicular Networks,'' {\it IEEE Trans. Veh. Technol.}, vol.~65, no.~6, pp.~3845--3859, June~2016.

\bibitem{DPalomar2006}
D. P. Palomar and Mung Chiang, ``A Tutorial on Decomposition Methods for Network Utility Maximization,''  {\it IEEE J. Sel. Areas  Commun.}, vol.~24, no.~8, pp.~1439--1451, Aug.~2006.

\bibitem{SZhang2016}
S. Zhang, N. Zhang, S. Zhou, J. Gong, Z. Niu, and X. Shen, ``Energy-Aware Traffic Offloading for Green Heterogeneous Networks,'' {\it IEEE J. Sel. Areas  Commun.}, vol.~34, no.~5, pp.~1116--1129, May~2016.

\bibitem{YWu2017TMC}
Y. Wu, J. Chen, L. Qian, J. Huang, and X. Shen, ``Energy-Aware Cooperative Traffic Offloading via Device-to-Device Cooperations: An Analytical Approach,'' {\it IEEE Trans. Mobile Comput.}, vol.~16, no.~1, pp.~97--114, Jan.~2017.

\bibitem{YZhou2016TWC}
Y. Zhou and W. Zhuang, ``Performance Analysis of Cooperative Communication in Decentralized Wireless Networks with Unsaturated Traffic,'' {\it IEEE Trans. Wireless Commun.}, vol.~15, no.~15, pp.~3518–-3530, Jan.~2016.

\bibitem{KMalekshan2017TWC}
K. Rahimi Malekshan and W. Zhuang, ``Joint Scheduling and Transmission Power Control in Wireless Ad Hoc Networks,'' {\it IEEE Trans. Wireless Commun.}, vol.~16, no.~9, pp.~5982--5993, June~2017. 

\bibitem{RVohra2011}
R. V. Vohra. Mechanism Design: A Linear Programming Approach. Cambridge University Press. New York, USA, 2011.

\bibitem{OSemiari2015}
O. Semiari, W. Saad, S. Valentin, M. Bennis, and H. V. Poor, ``Context-Aware Small Cell Networks: How Social Metrics Improve Wireless Resource Allocation,'' {\it IEEE Trans. Wireless Commun.}, vol.~14, no.~11, pp.~5927--5940, Nov.~2015.

\bibitem{SHe2017}
S. He, D. H. Shin, J. Zhang, J. Chen, and P. Lin, ``An Exchange Market Approach to Mobile Crowdsensing: Pricing, Task Allocation, and Walrasian Equilibrium,'' {\it IEEE J. Sel. Areas  Commun.}, vol.~35, no.~4, pp.~921--934, April~2017.

\bibitem{NLuong2017}
N. C. Luong, D. T. Hoang, P. Wang, D. Niyato, and Z. Han, ``Applications of Economic and Pricing Models for Wireless Network Security: A Survey,'' {\it IEEE Commun. Surveys Tuts.}, 2017, to appear.

\bibitem{NZhang2017}
N. Zhang, S. Zhang, J. Zheng, X. Fang, J.W. Mark, and X. Shen, ``QoE Driven Decentralized Spectrum Sharing in 5G Networks: Potential Game Approach'', {\it IEEE Trans. Veh. Technol}, vol.~66, no.~5, pp.~7797--7808, Mar.~2017. 

\bibitem{YN2009}
Y. Narahari, D. Garg, R. Narayanam, and H. Prakash. Game Theoretic Problems in Network Economics and Mechanism Design Solutions. {\it Advanced Information and Knowledge Processing Processing}. Springer-Verlag London. 1st~Edt., Feb.~2009. 

\bibitem{PSamadi2012}
P. Samadi, H. Mohsenian-Rad, R. Schober, and V. W. S. Wong, ``Advanced Demand Side Management for the Future Smart Grid Using Mechanism Design,'' {\it IEEE Trans. Smart Grid}, vol.~3, no.~3, pp.~1170--1180, Sept.~2012.

\bibitem{GIosifidis}
G. Iosifidis, L. Gao, J. Huang, and L. Tassiulas, ``A Double-Auction Mechanism for Mobile Data-Offloading Markets,'' {\it IEEE/ACM  Trans. Netw.}, vol.~23, no.~5, pp.~1634--1647, Oct.~2015.

\bibitem{TT2013}
T. Tanaka, F. Farokhi, and C. Langbort, ``A Faithful Distributed Implementation of Dual Decomposition and Average Consensus Algorithms,'' in {\it Proc. 52nd IEEE Conf. Decision and Control},  Dec.~2013, pp. 2985--2990, Firenze, Italy.

\bibitem{YWU2011}
Y. Wu, T. Zhang, and D.H.K. Tsang, ``Joint Pricing and Power Allocation for Dynamic Spectrum Access Networks with Stackelberg Game Model,'' {\it IEEE Trans. Wireless Commun.}, vol.~10,  no.~1, pp.~12--19, Jan. 2011.

\bibitem{AJin2017}
A.-L. Jin, W. Song, and W. Zhuang, ``Auction-based Resource Allocation for Sharing Cloudlets in Mobile Cloud Computing,'' {\it IEEE Trans. Emerg. Topics Comput.}, to appear.

\bibitem{ZZheng2017}
Z. Zheng, L. Song, and Z. Han, ``Bridge the Gap Between ADMM and Stackelberg Game: Incentive Mechanism Design for Big Data Networks,'' {\it IEEE Signal Process. Lett.}, vol.~24, no.~2, pp.~191--195, Feb.~2017.

\bibitem{HYaiche2000}
H. Yaiche, R. R. Mazumdar, and C. Rosenberg, ``A game theoretic framework for bandwidth allocation and pricing in broadband networks,'' {\it IEEE/ACM  Trans. Netw.}, vol.~8, no.~5, pp.~667--678, Oct.~2000.

\bibitem{XDuan2017}
X. Duan, C. Zhao, S. He, P. Cheng, and J. Zhang, ``Distributed Algorithms to Compute Walrasian Equilibrium in Mobile Crowdsensing,'' {\it IEEE Trans. Ind. Electron.}, vol.~64, no.~5, pp.~4048--4057, May~2017.

\bibitem{STSu2017}
S. T. Su, B. Y. Huang, C. Y. Wang, C. W. Yeh, and H. Y. Wei, ``Protocol Design and Game Theoretic Solutions for Device-to-Device Radio Resource Allocation,'' {\it IEEE Trans. Veh. Technol.}, vol.~66, no.~5, pp.~4271--4286, May~2017.

\bibitem{YZhao2017}
Y. Zhao and W. Song, ``Truthful Mechanisms for Message Dissemination via Device-to-Device Communications,'' {\it IEEE Trans. Veh. Technol.}, vol.~66, no.~11, pp.~10307--10321, Nov.~2017.

\bibitem{WSongGlobecom2016}
W. Song and Y. Zhao, ``A Randomized Reverse Auction for Cost-Constrained D2D Content Distribution,'' in {\it Proc. IEEE Globecom'16}, Washington, DC USA, Dec.~4--8, 2016.

\bibitem{PLi2016}
P. Li, S. Guo, and I. Stojmenovic, ``A Truthful Double Auction for Device-to-Device Communications in Cellular Networks,''  {\it IEEE J. Sel. Areas  Commun.}, vol.~34, no.~1, pp.~71--81, Jan.~2016.

\bibitem{CWang2016}
C. Y. Wang, G. Y. Lin, C. C. Chou, C. W. Yeh, and H. Y. Wei, ``Device-to-Device Communication in LTE-Advanced System: A Strategy-Proof Resource Exchange Framework,'' {\it IEEE Trans. Veh. Technol.}, vol.~65, no.~12, pp.~10022--10036, Dec.~2016.

\bibitem{CXu2013}
C. Xu, L. Song, Z. Han,  Q. Zhao,  X. Wang,  X. Cheng, and B. Jiao, ``Efficiency Resource Allocation for Device-to-Device Underlay Communication Systems: A Reverse Iterative Combinatorial Auction Based Approach,'' {\it IEEE J. Sel. Areas  Commun.}, vol.~31, no.~9, pp.~348--358, Sept.~2013.

\bibitem{LXu2016}
L. Xu, C. Jiang, Y. Shen, T. Q. S. Quek, Z. Han, and Y. Ren, ``Energy Efficient D2D Communications: A Perspective of Mechanism Design,'' {\it IEEE Trans. Wireless Commun.}, vol.~15, no.~11, pp.~7272--7285, Nov.~2016.

\bibitem{HKebriaei2016}
H. Kebriaei, B. Maham, and D. Niyato, ``Double-Sided Bandwidth-Auction Game for Cognitive Device-to-Device Communication in Cellular Networks,'' {\it IEEE Trans. Veh. Technol.}, vol.~65, no.~9, pp.~7476--7487, Sept.~2016.

\bibitem{AHajiesmaili2017}
M. H. Hajiesmaili, L. Deng, M. Chen, and Z. Li, ``Incentivizing Device-to-Device Load Balancing for Cellular Networks: An Online Auction Design,'' {\it IEEE J. Sel. Areas  Commun.}, vol.~35, no.~2, pp.~265--279, Feb.~2017.

\bibitem{RZhang2011}
R. Zhang, L. Song, Z. Han, and B. Jiao, ``Improve Physical Layer Security in Cooperative Wireless Network Using Distributed Auction Games,'' {\it in Proc. IEEE INFOCOM Workshops}, April~2011, Shanghai, China.

\bibitem{NNisan2007}
N. Nisan, T. Roughgarden, E. Tardos, and V. V. Vazirani. Algorithmic Game Theory. Cambridge University Press, 2007. New York, NY, USA.

\bibitem{CJiang2017}
C. Jiang, L. Kuang, Z. Han, Y. Ren, and L. Hanzo, ``Information Credibility Modeling in Cooperative Networks: Equilibrium and Mechanism Design,'' {\it IEEE J. Sel. Areas  Commun.}, vol.~35, no.~2, pp.~432--448, Feb.~2017.

\bibitem{TBorgers2015}
Tilman B\"{o}rgers. An Introduction to the Theory of Mechanism Design. Oxford University Press, 1 edition, 2015.

\bibitem{Namerikawa2015}
T. Namerikawa, N. Okubo, R. Sato, Y. Okawa, and M. Ono, ``Real-Time Pricing Mechanism for Electricity Market With Built-In Incentive for Participation,'' {\it IEEE Trans. Smart Grid}, vol.~6, no.~6, pp.~2714--2724, Nov.~2015.

\bibitem{KSingh2015}
K. Singh and M. L. Ku, ``Toward Green Power Allocation in Relay-Assisted Multiuser Networks: A Pricing-Based Approach,'' {\it IEEE Trans. Wireless Commun.}, vol.~14, no.~5, pp.~2470--2486, May~2015.


\bibitem{ELopez2016}
E. Egea-Lopez and P. Pavon-Mariño, ``Distributed and Fair Beaconing Rate Adaptation for Congestion Control in Vehicular Networks,'' {\it IEEE Trans. Mobile Comput.}, vol.~15, no.~12, pp.~3028--3041, Dec.~2016.

\bibitem{KShen2014}
K. Shen and W. Yu, ``Distributed Pricing-Based User Association for Downlink Heterogeneous Cellular Networks,'' {\it IEEE J. Sel. Areas  Commun.}, vol.~32, no.~6, pp.~1100--1113, June~2014.

\bibitem{NCLuong2017}
N. C. Luong, D. T. Hoang, P. Wang, D. Niyato and Z. Han, ``Applications of Economic and Pricing Models for Wireless Network Security: A Survey,'' {\it IEEE Commun. Surveys Tuts.}, to appear.

\bibitem{TR2014Lec}
T. Roughgarden. ``Demand Reduction in Multi-Unit Auctions Revisited''. Available at http://theory.stanford.edu/\texttildelow tim/w14/l/l37.pdf. Mar. 2014. 

\bibitem{LSong2014}
L. Song, D. Niyato, Z. Han, and E. Hossain, ``Game-theoretic resource allocation methods for device-to-device communication,'' in {\it IEEE Wireless Commun.}, vol.~21, no.~3, pp.~136--144, June~2014.

\bibitem{ZZhou2014}
Z. Zhou, M. Dong, K. Ota, J. Wu, and T. Sato, ``Energy Efficiency and Spectral Efficiency Tradeoff in Device-to-Device (D2D) Communications,'' {\it IEEE Wireless Commun. Lett.}, vol.~3, no.~5, pp.~485--488, Oct.~2014.

\bibitem{SSun2016}
S. Sun, T.~S.~Rappaport, S.~Rangan, T.~A.~Thomas, A.~Ghosh, I.~Z.~Kovacs, I.~Rodriguez, O.~Koymen, A.~Partyka, and J.~Jarvelainen, ``Propagation Path Loss Models for 5G Urban Micro- and Macro-Cellular Scenarios,'' in {\it Proc. IEEE Veh. Technol. Conf.}, May~2016, Nanjing, China.\\

%
%

%



\end{thebibliography}
\end{document}